\documentclass[11pt,a4paper]{article}
\usepackage{jheppub}
\usepackage[FIGTOPCAP]{subfigure}
\usepackage{wrapfig}
\usepackage{tensor}

\renewcommand{\Im}{\mathrm{Im}}
\renewcommand{\Re}{\mathrm{Re}}

\title{A strongly coupled anyon material}
\author[a,b]{Daniel K.~Brattan\note{E-mail address: danny.brattan@gmail.com}}
\affiliation[a]{Physics Department, Technion- Israel Institute of Technology,
Technion City - Haifa, \\ 32000, Israel.}
\affiliation[b]{Department of Mathematics-Physics-Computer Science, University
of Haifa at Oranim, \\ Qiryat Tivon, 36006, Israel.}

\abstract{We use alternative quantisation of the D3-D5 system to explore properties of a strongly coupled anyon material at finite density and temperature. We study the transport properties of the material and find both diffusion and massive holographic zero sound modes. By studying the anyon number conductivity we also find evidence for the anyonic analogue of the metal-insulator transition.}

\begin{document}

\maketitle
\flushbottom

%%%%%%%%%%%%%%%%%%%%%%%%%%%%%%%%%%%%%%%%%%%%%%%%%%%%%%%%%%%%%%%%%%%%%%%%%%%%%%%%%%%%%%%%%%%%%%%%%%%
%%%%%%%%%%%%%%%%%%%%%%%%%%%%%%%%%%%%%%%%%%%%%%%%%%%%%%%%%%%%%%%%%%%%%%%%%%%%%%%%%%%%%%%%%%%%%%%%%%%
%%%%%%%%%%%%%%%%%%%%%%%%%%%%%%%%%%%%%%%%%%%%%%%%%%%%%%%%%%%%%%%%%%%%%%%%%%%%%%%%%%%%%%%%%%%%%%%%%%%
\section{Introduction}
\label{background}
%%%%%%%%%%%%%%%%%%%%%%%%%%%%%%%%%%%%%%%%%%%%%%%%%%%%%%%%%%%%%%%%%%%%%%%%%%%%%%%%%%%%%%%%%%%%%%%%%%%
%%%%%%%%%%%%%%%%%%%%%%%%%%%%%%%%%%%%%%%%%%%%%%%%%%%%%%%%%%%%%%%%%%%%%%%%%%%%%%%%%%%%%%%%%%%%%%%%%%%
%%%%%%%%%%%%%%%%%%%%%%%%%%%%%%%%%%%%%%%%%%%%%%%%%%%%%%%%%%%%%%%%%%%%%%%%%%%%%%%%%%%%%%%%%%%%%%%%%%%

% AdSCMT programme
{\ A recent programme in the application of gauge/gravity duality to condensed matter systems is a move away from novel phenomena such as high temperature superconductivity towards reproducing well understood physics such as metal-insulator transitions. This approach is novel because the field theories in question are strongly coupled and quite distinct from the ``usual'' field theories of condensed matter. An example of standard condensed matter phenomenology that has been garnering interest in holography \cite{Jokela:2013hta,Brattan:2013wya,Jokela:2014wsa} is anyon physics in $(2+1)$-dimensions. We continue the study of \cite{Brattan:2013wya} which considers the low energy physics of a strongly coupled anyonic material in a magentic field by examining the same material in the absence of a magnetic field.}

% Anyon physics
{\ Anyons are particles whose statistics interpolate between Fermi-Dirac and Bose-Einstein. They can be formed from either bosons or fermions by attaching some fixed unit of magnetic flux to each fundamental unit of charge. When two such particles are exchanged the Aharanov-Bohm effect in $(2+1)$-dimensions contributes to the exchange phase angle $\theta$ an extra angle dependent on the amount of attached flux. As such the exchange phase angle generically becomes fractional. The resultant many body wave-functions then have fractional statistics \cite{PhysRevLett.49.957,Arovas:1985yb}.}

{\ In field theory one way to formalise this ``attaching of flux'' is to consider $SL(2,\mathbb{Z})$ transformations of the generating functionals of $U(1)$ conserved currents. Consider one such generating functional at non-zero charge density $q$, magnetic field $B$ and temperature $T$. Under an $ST^{K}$ operation \cite{Witten:2003ya,Leigh:2003ez,Herzog:2007ij,Brattan:2013wya} the charge density $d_{*}$ and magnetic field $B_{*}$ of the ``anyonised'' system are
  \begin{eqnarray}
   \label{Eq:Anyonising}
   d_{*} = - 2 \pi B \; , \qquad B_{*} = q - \frac{K}{2\pi} B \; . 
  \end{eqnarray}
We can then consider the generating functional of the $U(1)$ current whose background number density is $d_{*}$. If the ground state of the anyonised theory is chosen to have no magnetic field ($B_{*}=0$) then we say the theory describes anyons. This is because in terms of the original charge density and magnetic field, $q$ and $B$, a unit of charge density is accompanied by $\frac{K}{2\pi}$ units of magnetic flux. More generally acting with $SL(2,\mathbb{Z})$ transformations on the generating functionals for $U(1)$ currents creates whole families of theories and may be useful in understanding the quantum Hall effect. Given that such anyon particles have yet to be observed, but nonetheless do seem quite natural in $(2+1)$-dimensional systems, perhaps holography can help to identify key signatures for their existence.}

% Summary and paper layout
{\ In this paper we describe the low energy physics of a particular strongly coupled anyon material. We shall consider the field theory dual to the D3-D5 probe brane system at non-zero temperature, density and magnetic field as our ``original theory'' and anyonise it to one with vanishing magnetic field but non-zero density and temperature. Such a model is sufficiently simple to allow us to calculate the anyon diffusion constant and DC conductivities analytically but complex enough to display features such as a Drude-like peak in the AC conductivity.}

% Summary and paper layout 2
{\ In section \ref{model} we detail the holographic model under consideration. We shall work exclusively with massless anyons for simplicity. Subsequently, in section \ref{model:spectral}, we consider the pole structure of the theory with specific reference to the spectral functions. There will be qualitative agreement between the anyon spectral function at large statistical parameter and the spectral function of the original field theory at large charge densities and small magnetic fields. Namely we find a sound-like mode that becomes gapped as ground state parameters are tuned. The difference between the original theory and the anyonised theory however is the following: a gapped zero sound in the original theory is a consequence of non-zero magnetic fields \cite{Goykhman:2012vy,Brattan:2012nb} while in the anyon theory it is a consequence of statistics. Curiously we shall also find the same qualitative behaviour for small statistical parameter although the zero sound mode will be distinct from that 
found in the large parameter regime. In sections \ref{model:zerosound}, \ref{model:diffusion} and \ref{model:ACconductivity} we consider in more detail zero sound, diffusion and the conductivities respectively highlighting where observed features may be generic. We shall find in the AC conductivity evidence for the anyonic analogue of the metal-insulator 
transition.}

%%%%%%%%%%%%%%%%%%%%%%%%%%%%%%%%%%%%%%%%%%%%%%%%%%%%%%%%%%%%%%%%%%%%%%%%%%%%%%%%%%%%%%%%%%%%%%%%%%%
%%%%%%%%%%%%%%%%%%%%%%%%%%%%%%%%%%%%%%%%%%%%%%%%%%%%%%%%%%%%%%%%%%%%%%%%%%%%%%%%%%%%%%%%%%%%%%%%%%%
%%%%%%%%%%%%%%%%%%%%%%%%%%%%%%%%%%%%%%%%%%%%%%%%%%%%%%%%%%%%%%%%%%%%%%%%%%%%%%%%%%%%%%%%%%%%%%%%%%%
\section{Holographic model}
\label{model}
%%%%%%%%%%%%%%%%%%%%%%%%%%%%%%%%%%%%%%%%%%%%%%%%%%%%%%%%%%%%%%%%%%%%%%%%%%%%%%%%%%%%%%%%%%%%%%%%%%%
%%%%%%%%%%%%%%%%%%%%%%%%%%%%%%%%%%%%%%%%%%%%%%%%%%%%%%%%%%%%%%%%%%%%%%%%%%%%%%%%%%%%%%%%%%%%%%%%%%%
%%%%%%%%%%%%%%%%%%%%%%%%%%%%%%%%%%%%%%%%%%%%%%%%%%%%%%%%%%%%%%%%%%%%%%%%%%%%%%%%%%%%%%%%%%%%%%%%%%%

% The brane configuration - particularly masslessness
{\ We will use as our bulk theory a deformation of the D3-D5 brane system where the D5-branes are probes of the background given by $N_{c}$ D3-branes at finite temperature. The contribution of a D5-brane to the bulk action\footnote{We shall only consider one or a small number of probe D5-branes in the present paper to avoid them ``blowing-up'' to a D7-brane \cite{Myers:1999ps,Kristjansen:2012ny,Kristjansen:2013hma}.} is
  \begin{eqnarray}
   \label{Eq:D3D5probe}
   \mathcal{S}_{\mathrm{D5}}^{(0)} &=& - T_{D5} \int d^{6} \xi \sqrt{-\det\left(g + F \right)} \;
  \end{eqnarray}
where $\xi$ are the embedding coordinates, $T_{D5}$ the tension of the D5 brane and $F$ the $U(1)$ world-volume field strength. We have absorbed a factor of $2\pi \alpha'$ into the field strength compared to the usual definition and thus it is dimensionless. As the D5 brane is treated as a probe we neglect its back-reaction upon the bulk metric which we must specify. We take the metric $g$ to be
  \begin{eqnarray}
   ds^2 & = & g_{tt}(r) dt^2  + g_{11}(r) \left( dx^2+dy^2+dz^2\right) + g_{rr}(r) dr^2 + \ell^2 ds^2_{S^5} \; , \nonumber \\
	& = & -\frac{r^2}{\ell^2}f(r)dt^2+\frac{r^2}{\ell^2}\left(dx^2+dy^2+dz^2\right)+\frac{\ell^2}{r^2}\frac{dr^2}{f(r)} + \ell^2ds^2_{S^5} \; , 
	    \label{Eq:BackgroundMetric} \\
    f(r) & = & 1-\frac{r_H^4}{r^4} \; . \nonumber
  \end{eqnarray}
where $r=r_{H} = \pi T \ell^2$ with $\ell$ the AdS{} radius. This metric describes a thermal state which acts as a heat bath in the field theory to the fermions and bosons described by the probe brane. We now choose $\ell \equiv 1$. The embedding of the probe brane we will consider is the usual massless black hole embedding, maintaining chiral symmetry, with some background $U(1)$ baryon number charge carried by bosonic and fermionic excitations \cite{Karch:2002sh}. This model was considered in the anyon context in \cite{Brattan:2013wya}. As we have chosen to work with massless bosons and fermions while maintaining chiral symmetry the embedding is determined by the gauge field configuration since all the scalar profiles are trivial.}

\begin{table}[t!]
  \centering
  \label{tab:branembedding}
  \begin{tabular}{|c||cccccccccc|} \hline
			 & $t$ & $x$ & $y$ & $z$ & $X^1$ & $X^2$ & $X^3$ & $X^4$ & $X^5$ & $X^6$\\ \hline \hline
    $N_c$ \,\,\, D$3$ & $\times$ & $\times$ & $\times$ & $\times$ & & &  &  & & \\
    $N_f$ \,\,\, D$5$ & $\times$ & $\times$ & $\times$ & & $\times$ & $\times$ & $\times$ & & & \\ \hline
  \end{tabular}
  \caption{The embeddings of the D3 and D5 branes in ten dimensional Minkowski space.}
\end{table}

% Alternate quantisation
{\ The action of the anyonised D3-D5 probe brane model is given by
  \begin{eqnarray}
    \label{Eq:BackgroundAnyonAction}
    S^{(0)}_{\mathrm{anyon}} = S^{(0)}_{\mathrm{D5}} - \mathcal{N}_{5} \int d^{2+1}x \; \left[ A_{\mu}^{(0)}(x) A^{\mu}_{(1)}(x) + \frac{n}{2} \epsilon^{\mu \nu \rho} A_{\mu}^{(0)}(x) \partial_{\nu} A_{\rho}^{(0)}(x) \right] 
  \end{eqnarray}
where we work in $A_{r}(r,x^{\mu}) \equiv 0$ gauge, $\mathcal{N}_{5}=T_{\mathrm{D5}} V_{S^2}$ and 
  \begin{eqnarray}
   A_{M}(r,x) \mathrm{d} x^{M} &\stackrel{r \gg 1}{=}& \left[ A_{\mu}^{(0)}(x) + \frac{1}{r} A_{\mu}^{(1)}(x) + \mathcal{O}(1/r^2) \right] \mathrm{d} x^{\mu} \; .
  \end{eqnarray}
Here the $K$ of \eqref{Eq:Anyonising} is given by $2 \pi \mathcal{N}_{5} n$. See \cite{Brattan:2013wya} for further discussion of this action and more general $SL(2,\mathbb{Z})$ transformations. The interpretation of our bulk gauge field at the AdS boundary will be a little unusual in the holographic context so we shall now review it briefly. In the holographic dictionary we expand the bulk gauge field near the boundary and the leading and subleading terms provide data about some perturbing source and the subsequent expectation value of the operator to which it couples in the strongly coupled field theory. We parameterise these falloffs as
  \begin{eqnarray}
   \label{Eq:Alternatefalloffs}
%    b_{s} &=& 1 \; , c_{s} = - 1 \; , d_{s} = - K \;, n = \frac{d_{s}}{2 \pi \mathcal{N}_{5} c_{s}} = \frac{K}{2 \pi \mathcal{N}_{5}} \\
   A_{M}(r,x^{\mu}) \mathrm{d} x^{M} &=& \left[ v_{\mu}^{*}(x) + \frac{1}{r} \epsilon\indices{_{\mu}^{\nu \rho}} \partial_{\nu} \left( n v^{*}_{\rho}(x) + \frac{1}{2 \pi \mathcal{N}_{5}} A^{*}_{\rho}(x) \right) + \mathcal{O}(1/r^2) \right] \mathrm{d}x^{\mu} \; . 
  \end{eqnarray}
Subsequently, in the field theory, we interpret
  \begin{eqnarray}
   \label{Eq:AlternateCurrent}
   J^{\mu}_{*}(x) = \frac{1}{2 \pi} \epsilon^{\mu \nu \rho} \partial_{\nu} v_{\rho}^{*}(x)
  \end{eqnarray}
as the expectation value of the operator sourced by the external gauge field $A_{\mu}^{*} \mathrm{d} x^{\mu}$. We shall choose the Euclideanised time component of the gauge field $A_{\tau}^{*}$ to satisfy
  \begin{eqnarray}
   \label{Eq:ChemicalPotentialDef}
   \mu_{*} = i T \int_{0}^{\frac{1}{T}} d\tau \; A_{\tau}^{*} \; , 
  \end{eqnarray}
where $\mu_{*}$ is interpreted as the non-zero anyon number chemical potential. The ground states described by the action \eqref{Eq:BackgroundAnyonAction} are then those of massless thermal anyons with statistical parameter $n$ at non-zero anyon density.}

% The background solution
{\ Varying either the action of \eqref{Eq:D3D5probe} or \eqref{Eq:BackgroundAnyonAction} with respect to the bulk gauge field yields the same equation of motion as the additional term in \eqref{Eq:BackgroundAnyonAction} is a boundary term \cite{Marolf:2006nd}. We will demand that our ground state is stationary and preserves various symmetries: time translation, spatial translations and $SO(2)$ spatial rotation invariance. Given our restrictions it is straightforward to check that a solution to the bulk equations of motion is
  \begin{eqnarray}
   A_{t}(r) &=& c_{1} \left( \frac{1}{r} \;_{2}F_{1}\left[ \frac{1}{4}, \frac{1}{2}, \frac{5}{4}; - \frac{c_{1}^2+c_{2}^2}{r^4} \right]
				 - \frac{1}{\pi T} \;_{2}F_{1}\left[ \frac{1}{4}, \frac{1}{2}, \frac{5}{4}; - \frac{c_{1}^2+c_{2}^2}{(\pi T)^4} \right] \right) \; ,\\
   A_{2}(x^{1}) &=& c_{2} x^{1} \; ,
  \end{eqnarray}
where $i,j$ denote $x^{1}, x^{2}$, $c_{1}$ and $c_{2}$ are arbitrary constants and we have used regularity at the future horizon. Expanding near the boundary we identify
  \begin{eqnarray}
    v_{t}^{*} = - \frac{c_{1}}{\pi T} \;_{2}F_{1}\left[ \frac{1}{4}, \frac{1}{2}, \frac{5}{4}; - \frac{c_{1}^2+c_{2}^2}{(\pi T)^4} \right] \; , \qquad
    v_{1}^{*} = 0 \; , \qquad
    v_{2}^{*} = c_{2} x^{1} \; , 
  \end{eqnarray}
from the leading terms and from the subleading terms
  \begin{eqnarray}
       \epsilon^{ij} \partial_{j} \left( n v_{t}^{*} + \frac{1}{2 \pi \mathcal{N}_{5}} A_{t}^{*} \right) = 0 \; , \qquad
   c_{1} = n \left( \partial_{1} v_{2}^{*} - \partial_{2} v_{1}^{*} \right) + \frac{1}{2 \pi \mathcal{N}_{5}} \left( \partial_{1} A_{2}^{*} - \partial_{2} A_{1}^{*} \right) \; ,
  \end{eqnarray}
where we have used stationarity to simplify the first constraint. Imposing \eqref{Eq:ChemicalPotentialDef} for constant $\mu_{*}$ then satisfies the first constraint while choosing zero magnetic field in the ground state, $B_{*}=0$, means that 
  \begin{eqnarray}
   c_{1} &=& n \left( \partial_{1} v_{2}^{*} - \partial_{2} v_{1}^{*} \right) \; .
  \end{eqnarray}
Identifying $-\frac{c_{1}}{2 \pi n} = d_{*}$, with $d_{*}$ the anyon number density, then additionally implies that $c_{2}= - 2\pi d_{*}$ completely fixing our ground state solution to be
  \begin{eqnarray}
   A_{t}(r) &=& - \frac{2 \pi n d_{*}}{r} \;_{2}F_{1}\left[ \frac{1}{4}, \frac{1}{2}, \frac{5}{4}; - (2 \pi d_{*})^2 \left( \frac{n^2 + 1}{r^4} \right) \right] \nonumber \\
	    &\;& + \frac{2 n d_{*}}{T} \;_{2}F_{1}\left[ \frac{1}{4}, \frac{1}{2}, \frac{5}{4}; - (2 \pi d_{*})^2 \left( \frac{n^2 + 1}{(\pi T)^4} \right) \right] \; , \qquad \\
   A_{1}(x^{2}) &=& 0 \; , \qquad A_{2}(x^{1}) = - 2 \pi d_{*} x^{1} \; .
  \end{eqnarray}
This ground state also solves the equations of motion coming from \eqref{Eq:BackgroundAnyonAction}. Again this is because the additional term in \eqref{Eq:BackgroundAnyonAction} is a boundary term. In terms of the theory described by \eqref{Eq:D3D5probe} the constants $c_{1}$ and $c_{2}$ are the boundary charge density $d=q/\mathcal{N}_{5}$ and the boundary magnetic field $B$. The relation between $c_{1}$ and $c_{2}$ can then be written as $d=nB$ and we see that in terms of the original theory this anyonised theory describes particles with $\frac{K}{2\pi} = n \mathcal{N}_{5}$ units of magnetic flux attached to each unit charge (compare with \eqref{Eq:Anyonising}).}

% The equations of motion for the fluctuations
{\ Now we derive the equations of motion for density fluctuations about our thermal, non-zero anyon density ground state. Again because the difference between \eqref{Eq:D3D5probe} and \eqref{Eq:BackgroundAnyonAction} is a boundary term the equations of motion for the fluctuations are the same. We shall allow the perturbations of the boundary field theory to be space and time dependent. However we can make use of spatial $SO(2)$ rotation invariance in the boundary and restrict to spatial dependence in the $x^{1}$ direction. Upon replacing $A(r,x^{\mu}) \mapsto A(r,x^{\mu}) + a(r,x^{\mu})$ and Fourier decomposing
  \begin{eqnarray}
   a_{\mu}(r,x) &=& \int \frac{d\omega dk_{1}}{(2 \pi)^2} a_{\mu}(r,k_{1}) \exp \left( - i \omega t + i k_{1} x^{1} \right) \; ,
  \end{eqnarray}
one can obtain the fluctuation equations of motion. These equations overspecify the problem due to gauge redundancy in the bulk and are thus unilluminating. As such we only record the $a_{r}$ equation which translates to a gauge constraint
  \begin{eqnarray}
    \label{eq:areom}
    \omega \, a_t' + u(r)^2 k_{1} \, a_1' =0,
  \end{eqnarray}
where
  \begin{eqnarray}
    u(r)^2 \equiv \frac{|g_{tt}|g_{rr} - A_t'^2}{\frac{g_{rr}}{g_{11}} \left(g_{11}^2 + \left(2 \pi d_{*}\right)^2\right)} = \frac{|g_{tt}|g_{11}}{g_{11}^2 + \left( 1 + n^2 \right) \left(2 \pi d_{*} \right)^2} \; , 
  \end{eqnarray}
and prime denotes differentiation with respect to $r$. By aligning the momentum along the $x^{1}$-direction we find that the gauge-invariant fluctuations are $a_2$ and
  \begin{eqnarray}
    E_{1}(r,\omega,k_{1}) \equiv k_{1} \, a_t(r,\omega,k_{1}) + \omega \, a_1(r,\omega,k_{1}) .
  \end{eqnarray}
The equations of motion for the gauge fluctuations in terms of $E_{1}$ and $a_{2}$ are then {\small
  \begin{subequations}
    \begin{eqnarray}
      \label{eq:Eeom}
    E_{1}'' &+& \left [ \partial_r \log \left( \frac{g_{11}^{3/2}|g_{tt}| g_{rr}^{-1/2}}{\left(\omega^2 - u(r)^2 k_{1}^2\right)u(r)\left(g_{11}^2 + \left(2 \pi d_{*}\right)^2 \right)} \right)\right] E_{1}' + \frac{g_{rr}}{|g_{tt}|}\left(\omega^2 - u(r)^2 k_{1}^2\right) E_{1} \qquad \nonumber \\
	  &=& - i \left(2 \pi d_{*}\right) \left[\frac{u(r)\left(g_{11}^2 + \left(2 \pi d_{*}\right)^2\right)g_{rr}^{1/2}}{|g_{tt}|g_{11}^{3/2}}\right] \left [ \partial_r \left(\frac{g_{11}^{1/2} g_{rr}^{-1/2} A_t'}{u(r)\left(g_{11}^2 + \left(2 \pi d_{*} \right)^2 \right)}\right)\right] \times \nonumber \\
	  & & \left(\omega^2 - u(r)^2 k_{1}^2\right) a_2 \; , \\
      \label{eq:ayeom}
    a_2'' &+& \left [ \partial_r \log \left(\frac{g_{11}^{3/2}|g_{tt}|g_{rr}^{-1/2}}{u(r)\left(g_{11}^2 + \left(2 \pi d_{*}\right)^2\right)}\right) \right] a_2'  + \frac{g_{rr}}{|g_{tt}|} \left(\omega^2 - u(r)^2 k_{1}^2\right)a_2 \nonumber \\
	  &=& + i \left(2 \pi d_{*}\right) \left[ \frac{u(r) \left(g_{11}^2 + \left(2 \pi d_{*}\right)^2\right) g_{rr}^{1/2}}{|g_{tt}|g_{11}^{3/2}}\right]\left[\partial_r \left(\frac{g_{11}^{1/2}g_{rr}^{-1/2}A_t'}{u(r)\left(g_{11}^2 + \left(2 \pi d_{*}\right)^2\right)}\right)\right] E_{1} \; .
    \end{eqnarray}
  \end{subequations}
}}

{\ We now layout the numerical procedure \cite{Kaminski:2009dh} used to determine solutions to the bulk equations, \eqref{eq:Eeom} and \eqref{eq:ayeom}, while satisfying the mixed quantisation condition. We shall employ the notation of \cite{Brattan:2012nb}. Eqs.~\eqref{eq:Eeom} and~\eqref{eq:ayeom} are second-order, hence for each field, $E_{1}$ and $a_2$, we need two boundary conditions to specify a solution completely. On the black hole horizon, a solution for $E_{1}$ or $a_{2}$ looks like a
linear combination of in-going and out-going waves, with some normalizations. The prescription for obtaining the \textit{retarded} Green's function requires that we choose our normalisations to remove any outgoing modes~\cite{Son:2002sd,Policastro:2002se,Skenderis:2008dh,vanRees:2009rw}. Let 
  \begin{eqnarray}
      \vec{V}(r,\omega,k_{1})\equiv\begin{pmatrix} E_{1}(r,\omega,k_{1}) \\ a_2(r,\omega,k_{1}) \end{pmatrix}
  \end{eqnarray}
and at large $r$ identify
  \begin{eqnarray}
   \vec{V}(r,\omega,k_{1}) = \vec{V}^{(0)}(\omega,k_{1}) + \frac{1}{r} \vec{V}^{(1)}(\omega,k_{1}) + \mathcal{O}(1/r^2) \; .
  \end{eqnarray}
For mixed quantisation a  boundary condition is given by fixing,
  \begin{eqnarray}
    \mathcal{N}_{5} \left[ \left( \begin{array}{cc} 1/p^2 & 0 \\ 0 & 1 \end{array} \right) \vec{V}^{(1)}(\omega,k_{1}) + i n \left( \begin{array}{cc} 0 & 1 \\ 1 & 0 \end{array} \right) \vec{V}^{(0)}(\omega,k_{1}) \right] = \vec{V}_{\mathrm{b}}=fixed
  \end{eqnarray}
to some value, denoted $\vec{V}_{\mathrm{b}}$, at the boundary. For numerical purposes however it is preferable to fix all our boundary conditions at the future black hole horizon. The second boundary condition is then the normalisation of the ingoing wave at the horizon. The two ways to fix boundary conditions are related to each other by a change of basis transformation. The vector of near-horizon normalization factors, $\vec{V}_{\mathrm{nh}}$, is, when the temperature is non-zero,
  \begin{eqnarray}
    \label{eq:vnhdef}
    \vec{V}_{\mathrm{nh}} \equiv \lim_{r \rightarrow r_{H}} \, \exp \left( i \omega \int dr \sqrt{g_{rr}/|g_{tt}|} \right) \vec{V}(r,\omega,k_{1}).
  \end{eqnarray}
Notice that $\vec{V}_{\mathrm{nh}}$ is constant, independent of $r$, $\omega$, and $k_{1}$. On the right-hand-side of eq.~\eqref{eq:vnhdef}, the exponential factor is designed to \textit{cancel} the exponential factor that represents an
in-going wave at the future horizon.}

{\ We now pick two convenient values of $\vec{V}_{\mathrm{nh}}$ and solve the equations for each of these choices. This provides us with a basis of solutions in terms of which we can write any solution. The typical choices we have used in our numerics are: $\vec{V}^{(1)}_{\mathrm{nh}}=(1,1)^T$ and $\vec{V}^{(2)}_{\mathrm{nh}}=(1,-1)^T$. Let us call the corresponding solutions $\vec{V}^{(1)}$ and $\vec{V}^{(2)}$ and use them to define a matrix $P(r,\omega,k_{1})$ by
 \begin{eqnarray}
    \label{eq:pmatrixdef}
    P(r,\omega,k_{1}) \equiv \left(\vec{V}^{(1)}(r,\omega,k_{1}),\vec{V}^{(2)}(r,\omega,k_{1})\right) \; .
 \end{eqnarray}
Using this matrix we can write any solution to the equations of motion with initial condition $\vec{V}_{\mathrm{nh}}$ at the horizon as
  \begin{eqnarray}
    \label{Eq:Diffproblem} 
    \vec{V}(r,p) = P(r,p) \, \vec{V}_{\mathrm{nh}}.
  \end{eqnarray}
In terms of the bulk to boundary propagator $P$ and the near horizon vector $\vec{V}_{\mathrm{nh}}$ we have
  \begin{eqnarray}
       \vec{V}_{\mathrm{b}}
   &=&  \mathcal{N}_{5} \lim_{r \rightarrow \infty} \left[ \left( \begin{array}{cc} 1/p^2 & 0 \\ 0 & 1 \end{array} \right) \left( -r^2 P'(r,\omega,k_{1}) \right) + i n \left( \begin{array}{cc} 0 & 1 \\ 1 & 0 \end{array} \right) P(r,\omega,k_{1}) \right]    
       \vec{V}_{\mathrm{nh}} \; . \nonumber
  \end{eqnarray}
We call any solution to the bulk equations with $\vec{V}_{\mathrm{b}}=0$ and complex frequency or complex momentum a quasi-normal mode. For a non-trivial solution with $\vec{V}_{\mathrm{b}}  \equiv 0$ and $\vec{V}_{\mathrm{nh}} \neq 0$ it must be the case that \cite{Kaminski:2009dh}
  \begin{eqnarray}
    \label{Eq:QNMCondition}
    \lim_{r \rightarrow \infty} \det \left[ \left( \begin{array}{cc} 0 & 1 \\ 1/p^2 & 0 \end{array} \right)  \left( -r^2 P'(r,\omega,k_{1}) \right) + i n P(r,\omega,k_{1}) \right] = 0
  \end{eqnarray}
which places a constraint on $\omega$ and $k_{1}$ yielding the dispersion relation of the mode\footnote{Note that the constraint for finding quasi-normal modes, as we have written it, is blind to poles at zero momentum and frequency.}. To satisfy this constraint it is generally necessary to complexify one of $\omega$ or $k_{1}$ after which we obtain dispersion relations of the form $\omega(k_{1})$ and $k_{1}(\omega)$ respectively. We shall have cause to use both and the choice of complexification is demonstrated by which variable is chosen to be a function of the other. The retarded Green's function can also be obtained from $P(r,\omega,k_{1})$ and is given by
  \begin{eqnarray}	
     G_{R}^{*}(p) &=& \frac{1}{(2 \pi)^2 \mathcal{N}_{5}} \lim_{r \rightarrow \infty} \left\{ \left( \begin{array}{cc} 0 & -\omega \\ 1 & 0 \end{array} \right) P(r,p) \times \right. \nonumber \\
	      &\;& \left. \left[ \left( \begin{array}{cc} 1/p^2 & 0 \\ 0 & 1 \end{array} \right) \left( - r^2 P'(r,p) \right) + i n \left( \begin{array}{cc} 0 & 1 \\ 1 & 0 \end{array} \right) P(r,p) \right]^{-1}     
		    \left( \begin{array}{cc} 0 & 1 \\ \omega & 0 \end{array} \right) \right\} \nonumber \\
	      &=& \left( \begin{array}{cc}
			             \langle J^{*}_{x}(p) J^{*}_{x}(-p) \rangle  & \langle J^{*}_{x}(p) J^{*}_{y}(-p) \rangle  \\
			             \langle J^{*}_{y}(p) J^{*}_{x}(-p) \rangle  & \langle J^{*}_{y}(p) J^{*}_{y}(-p) \rangle  \\
			            \end{array} \right) \; .
  \end{eqnarray}
If the reader would prefer components with $t$ as opposed to $x$ they need only replace the explicit factors of $\omega$ by $-k_{1}$.}

{\ Finally we note that because of scale covariance we can remove the temperature as a variable by scaling other variables, such as the anyon number density, by appropriate powers of $\pi T$. As such any object that comes with a tilde is invariant under a Weyl rescaling of the boundary metric.}

\begin{figure}[t]
 \centering
 \begin{subfigure}
  \centering
  \includegraphics[width=0.35\textwidth]{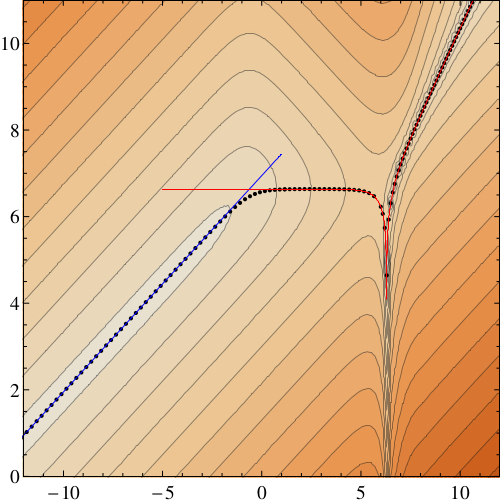}
 \end{subfigure} \qquad
 \begin{subfigure}
  \centering
  \includegraphics[width=0.35\textwidth]{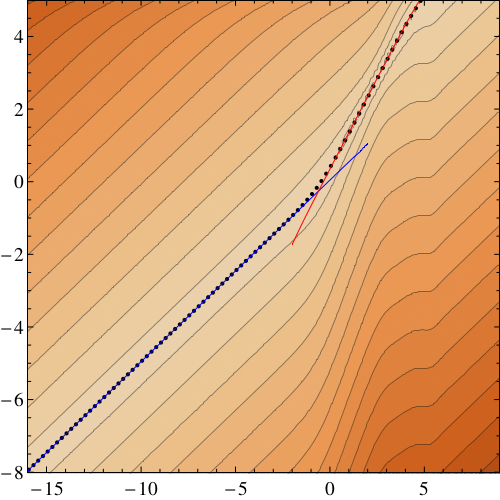}
 \end{subfigure} \\
 \begin{subfigure}
  \centering
  \includegraphics[width=0.35\textwidth]{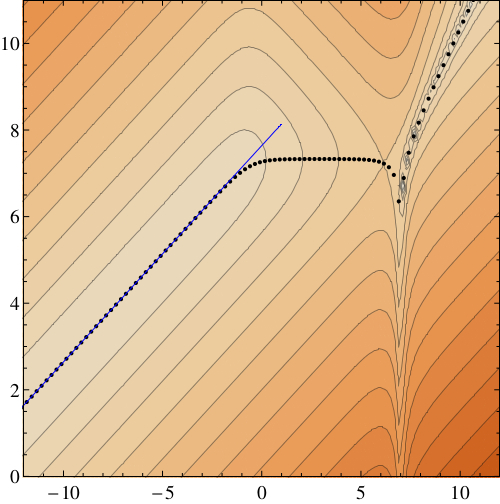}
 \end{subfigure} \qquad
 \begin{subfigure}
  \centering
  \includegraphics[width=0.35\textwidth]{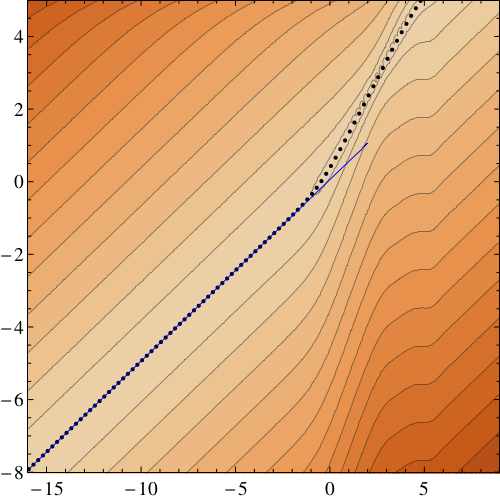}
 \end{subfigure} \\
 \begin{picture}(100,0)
  \put(80,1){$\log (\tilde{\omega} \sqrt{(2 \pi n) |\tilde{d}_{*}|})$}
  \put(3,84){$\log (\tilde{k} \sqrt{(2 \pi n) |\tilde{d}_{*}|})$}
 \end{picture}
 \caption{Properties of the logarithm of the time component of the spectral function, $\log \chi^{tt}$, for the D3-D5 system and its weakly anyonic counterpart against suitably normalised logarithms of frequency and momentum. The unusual axis normalisations are chosen to make comparison with the figures of \cite{Brattan:2012nb} simple. \textbf{Top:} The time component of the spectral function of the D3-D5 system at non-zero temperature $T$, charge $Q/(\pi V T) = 10^6 \mathcal{N}_{5}$ and magnetic field $B/(\pi T)^2=10^3$ (left) where $V$ is the regulating spatial volume at the boundary. The right hand uppermost figure has $Q/(\pi V T) = 10^2 \mathcal{N}_{5}$ and $B/(\pi T)^2=10^{-1}$ (right). The black dots represent the absolute value of quasi-normal mode closest to the real momentum axis, $|k_{1}(\omega)|$, and show a change from massive zero sound to diffusion. The solid blue and red lines represent the analytic expressions for the diffusive and zero sound modes given in \cite{Brattan:2012nb}. \textbf{Bottom:} The time 
component of the spectral function of the anyon system at non-zero temperature $T$, anyon density 
$\tilde{d}_{*}=-10^3/(2 \pi)$ and statistical parameter $n=10^3$ (left) and $\tilde{d}_{*}=-10^{-1}/(2 \pi)$ and $n=10^3$ (right). The solid blue line is given by the analytic expression for the diffusion constant in \eqref{Eq:EinsteinRelation}. We call the system weakly anyonised because of the qualitative similarity of the top and bottom figures ($n$ is relatively large). We see approximately the same behaviour for the weakly anyonised system as the D3-D5 probe brane system - namely a diffusive pole transitioning into a massive sound-like pole as the density increases.}
 \label{fig:Weakly_anyonised}
\end{figure}
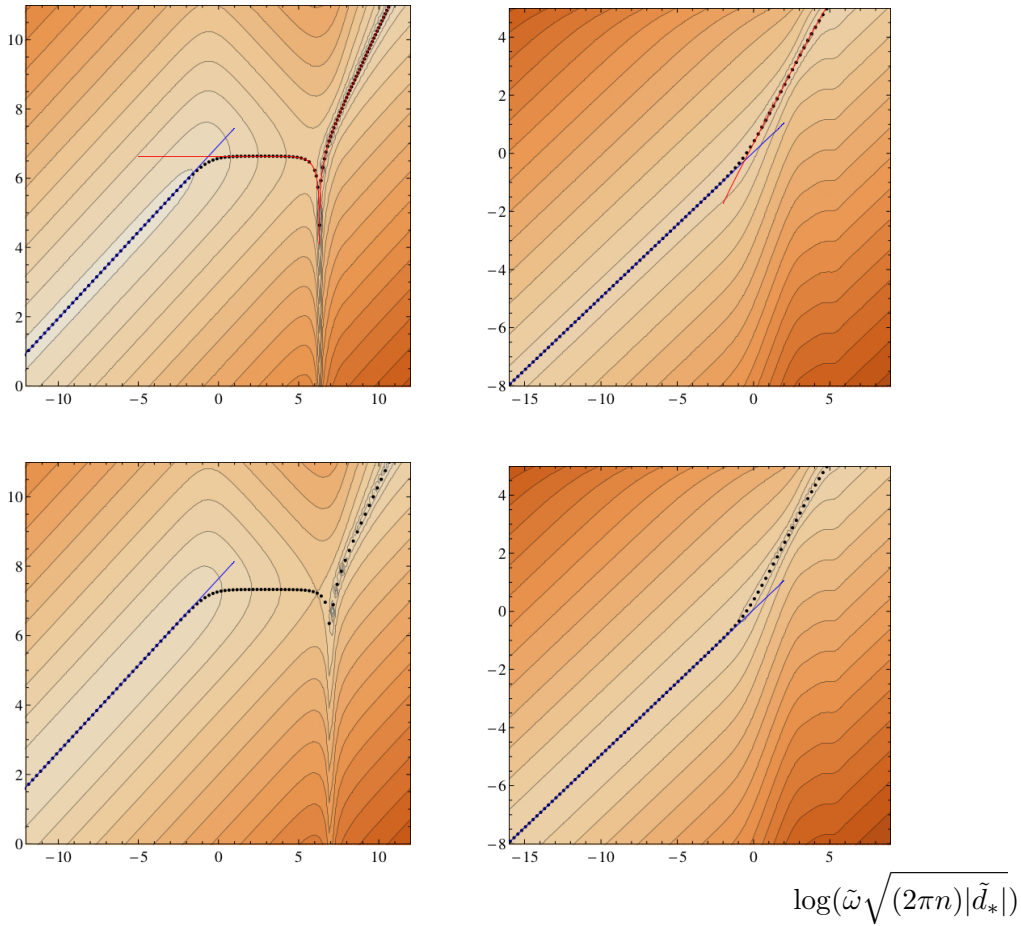

%%%%%%%%%%%%%%%%%%%%%%%%%%%%%%%%%%%%%%%%%%%%%%%%%%%%%%%%%%%%%%%%%%%%%%%%%%%%%%%%%%%%%%%%%%%%%%%%%%%
\subsection{The spectral function and poles}
\label{model:spectral}
%%%%%%%%%%%%%%%%%%%%%%%%%%%%%%%%%%%%%%%%%%%%%%%%%%%%%%%%%%%%%%%%%%%%%%%%%%%%%%%%%%%%%%%%%%%%%%%%%%%

% Definition of spectral function
{\ We now turn to a quasi-normal mode analysis to determine how the system relaxes perturbations. There are two regimes we shall be interested in, namely, late times and low frequencies and momenta with respect to the temperature. The late time behaviour can easily be understood by searching for poles with small imaginary parts in the complex frequency plane. These near origin poles also contribute in the low frequency and momentum regime. However the response of the system in this latter case can be strongly affected by the size of the residues of the Green's function at the poles. To complement the pole structure analysis therefore we also compute the spectral function
  \begin{eqnarray}
    \label{eq:spectralfuncdef}
    \chi^{\mu \nu}(\omega,k_{1}) \equiv i \left[ \left(G^{*}_R\right)^{\mu \nu}(\omega,k_{1})-\left(G^{*}_R\right)^{\mu \nu}(\omega,k_{1})^{\dagger}\right],
  \end{eqnarray}
where $\left(G^{*}_R\right)^{\mu \nu}(\omega,k_{1})$ is the retarded Green's function. The spectral function is a real-valued function of real $\omega$ and $k_{1}$, and hence is observable.  Physically, $\chi^{\mu \nu}(\omega,k_{1})$ determines the rate of work done on the system by a small external source \cite{Hartnoll:2009sz}. A pole in Green's function at complex frequency or momentum with sufficiently large residue will produce a large peak in $\chi^{\mu \nu}(\omega,k_{1})$. Given knowledge of the spectral function, which is the imaginary part of the Green's function we can reconstruct $\textrm{Re} \, G_R^{\mu \nu}(\omega,k_{1})$ using the Kramers-Kronig relation, provided the large-$\omega$ and large-$k_{1}$ asymptotics have been suitably regulated~\cite{Hartnoll:2009sz}.}

% The D3-D5 probe brane system at zero magnetic field
{\ We begin by considering a weakly anyonic ($n \gg 1$) system. The field theory consists of bosons and fermions with a small extra statistical angle and thus a naive expectation would be that the anyonised system closely resembles the original system. Hence we should briefly review the physics of the D3-D5 probe brane at non-zero temperature, density and magnetic field. It is already well known that the strongly coupled theory described by the probe branes has zero sound and diffusion modes \cite{Goykhman:2012vy,Gorsky:2012gi,Wapler:2009tr,Pal:2012gr,Brattan:2012nb}. Extending the Green's function to complex frequency and real momentum the diffusion mode is represented by a purely imaginary pole that starts at the origin as momentum goes to zero. As the momentum is increased it sinks to larger imaginary frequency until it connects with another imaginary pole from lower in the complex frequency plane. The result is two complex poles with opposite sign real parts i.e~two imaginary poles join to produce two 
poles with real parts (see \cite{Brattan:2012nb} for relevant diagrams). If the density is sufficiently large (see \cite{Brattan:2012nb} for the meaning of sufficiently large) then the complex pole will have the dispersion relation associated with zero sound. When it is too small the resultant complex pole has a dispersion relation whose real part has a gradient closer to one.}

% The D3-D5 probe brane system at non-zero magnetic field
{\ It was observed that for small magnetic field in D3-D5 and D3-D7 probe brane systems \cite{Brattan:2012nb} the qualitative motion of the poles at zero magnetic field continues to hold even though the zero sound mode becomes gapped \cite{Goykhman:2012vy,Gorsky:2012gi,Brattan:2012nb}. This is because the thermal energy of the background is sufficiently high so as to overcome the gap. However as the magnetic field is increased further the zero sound and diffusion poles become distinct and a region of reduced weight appears in the spectral function (see fig.~\ref{fig:Weakly_anyonised}). In this latter case the diffusion pole simply sinks deeper into the complex plane rather than connecting with another pole. Additionally at zero frequency the sound mode is already present as a complex pole. This complex pole has a smaller imaginary part than the diffusion pole for sufficiently large momentum and thus dominates the low energy excitations of the system. It should be noted that if we choose to make momentum 
complex in our Green's function then the pole motion is very 
different and diffusion and zero sound always come from the same single complex pole - no matter the strength of the magnetic field. It is the absolute value of this complex momentum pole that is shown by the black dots in fig.~\ref{fig:Weakly_anyonised}.}

% Weakly anyonised system
{\ Returning now to the anyon system we see precisely the same qualitative behaviour at large $n$. The lower spectral functions in fig.~\ref{fig:Weakly_anyonised} are very similar to those above although the peaks are displaced and in some cases broadened. It is important to emphasise however a difference between the original and anyonised systems. The former only develops a massive zero sound in the presence of a non-zero magnetic field. The latter has no magnetic field and develops a gap as a function of statistics alone. Additionally whether the zero sound gap is overcome by the thermal energy or not is purely dependent on the statistical parameter $n$. We shall compute the mass of the zero sound mode in section \ref{model:zerosound}.}

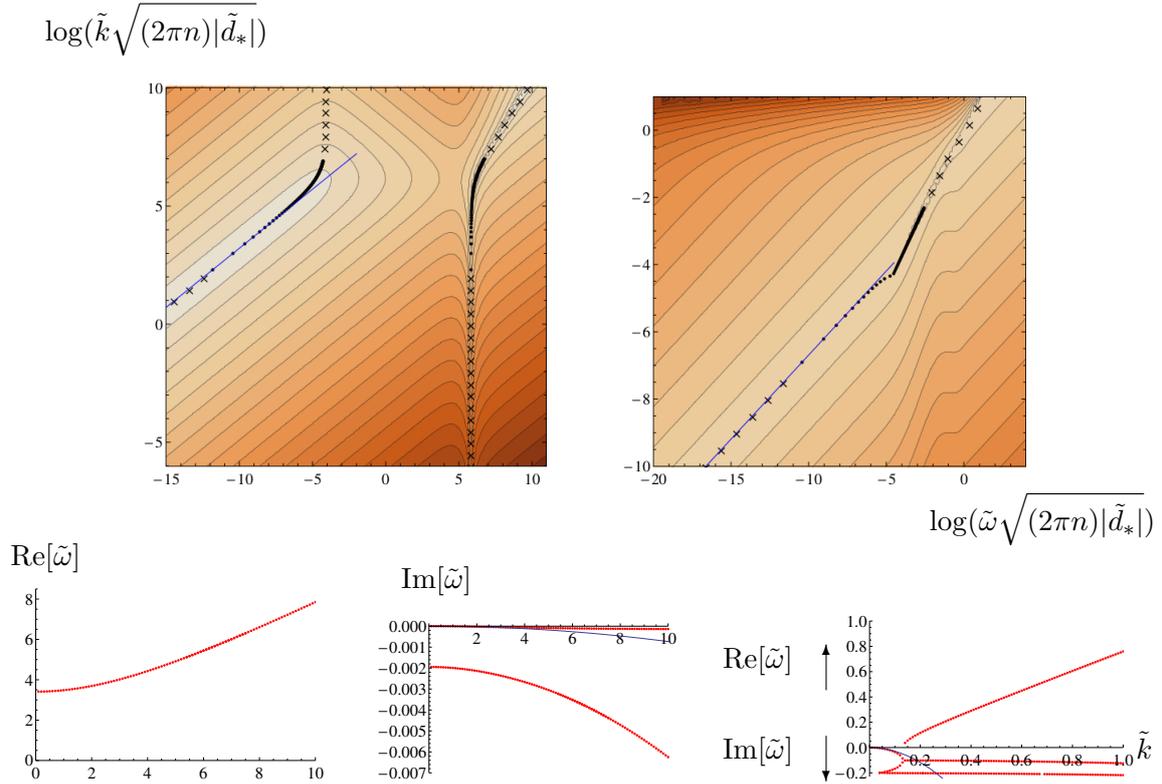
\begin{figure}[!t]
 \centering
 \begin{subfigure}
  \centering
  \includegraphics[width=0.35\textwidth]{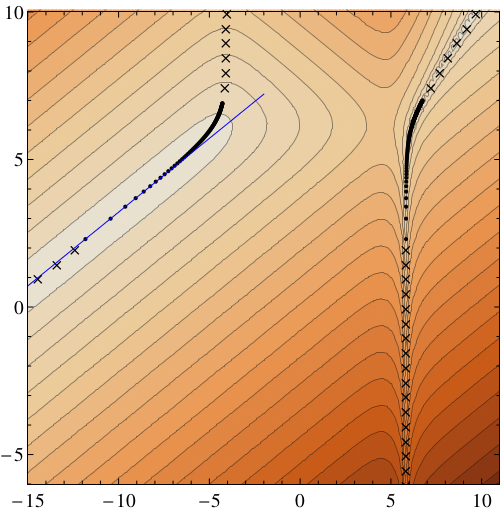}
 \end{subfigure} \qquad
 \begin{subfigure}
  \centering
  \includegraphics[width=0.35\textwidth]{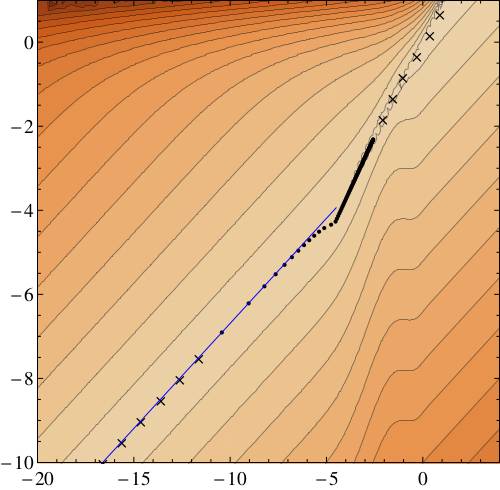}
 \end{subfigure} \\ \vskip+4\unitlength \hskip -1\unitlength
 \begin{subfigure}
  \centering
  \includegraphics[width=0.26\textwidth]{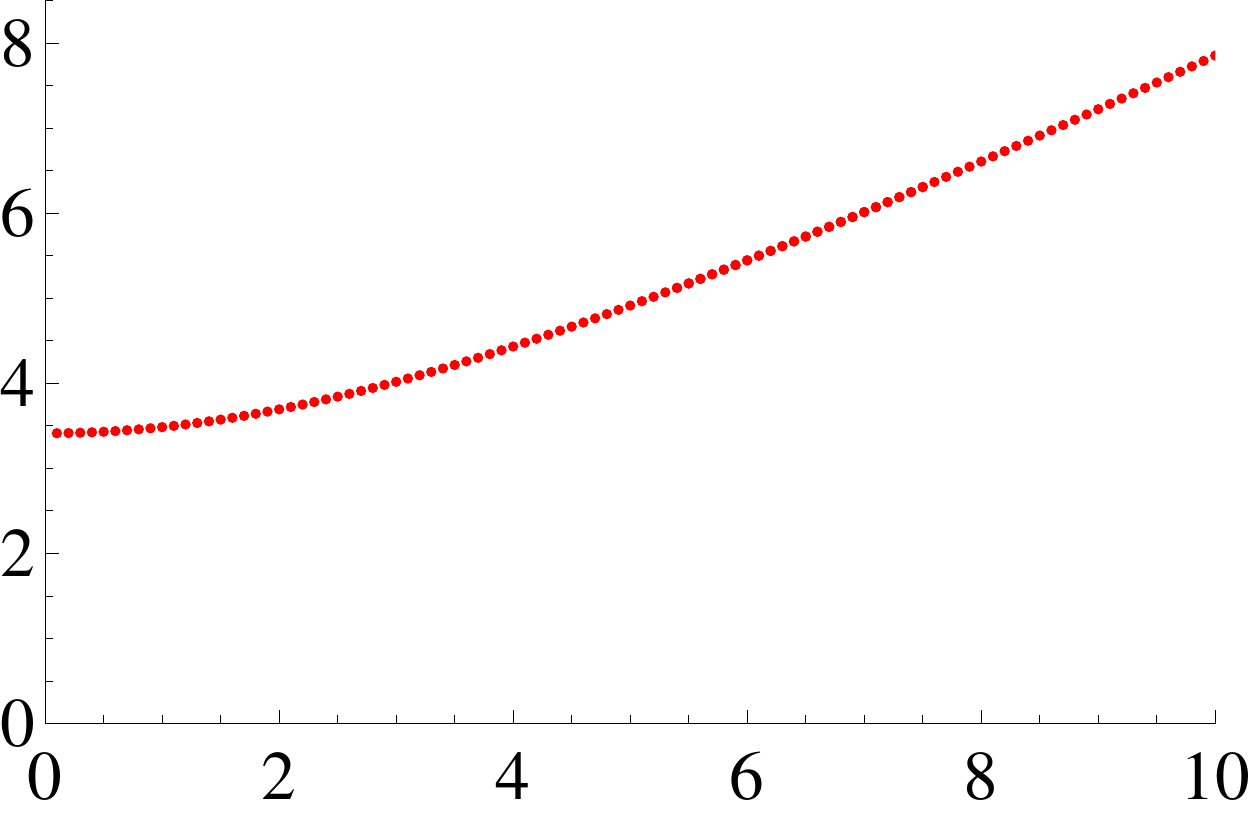}
 \end{subfigure} \hskip+3\unitlength 
 \begin{subfigure}
  \centering
  \includegraphics[width=0.26\textwidth]{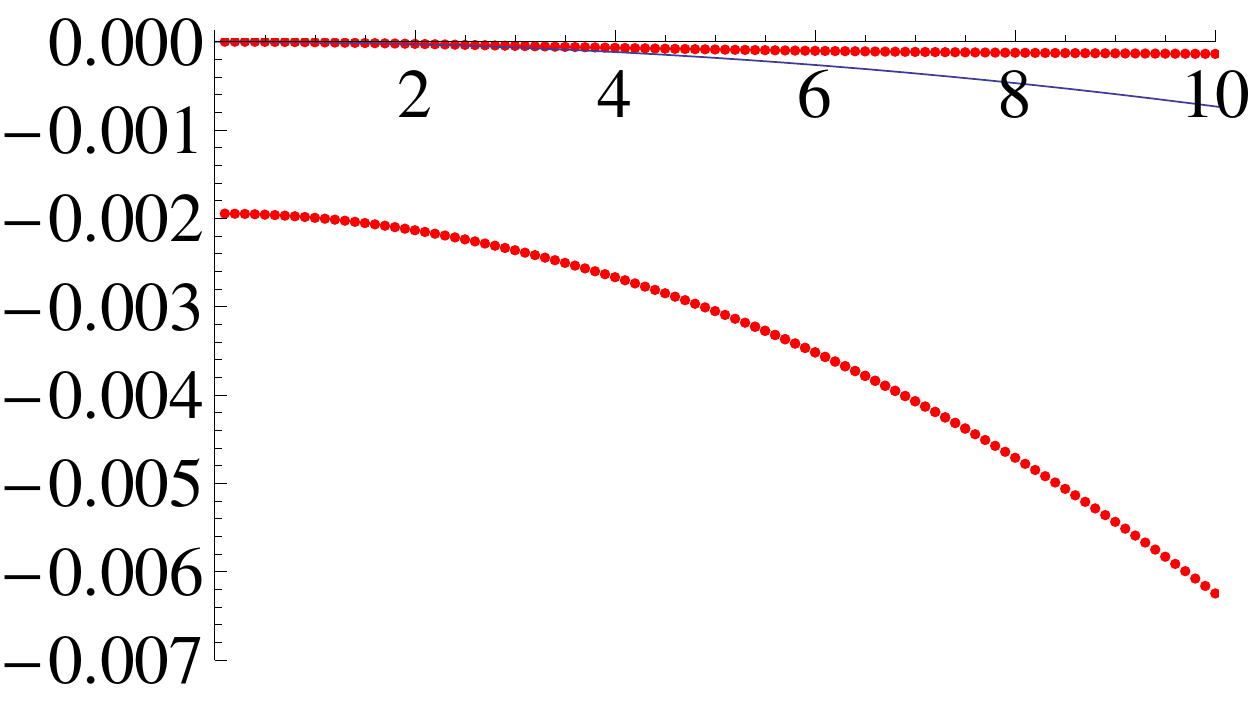}
 \end{subfigure} \qquad \hskip+7\unitlength
 \begin{subfigure}
  \centering
  \includegraphics[width=0.26\textwidth]{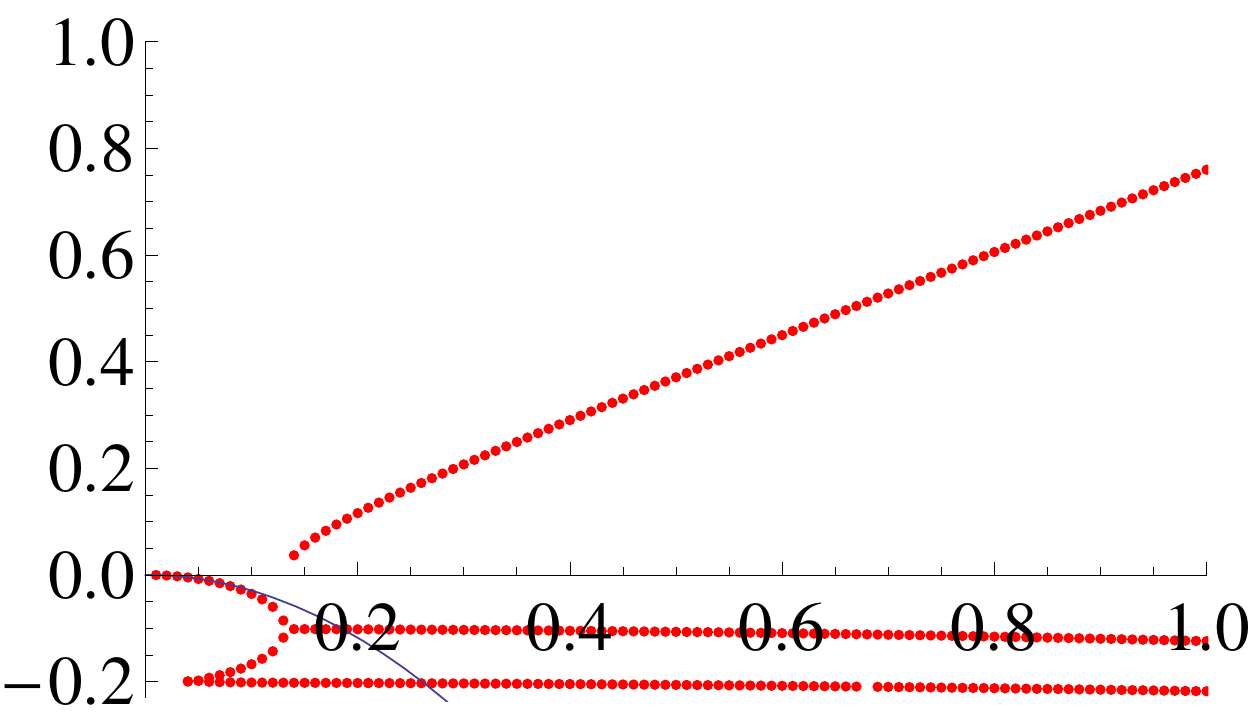}
 \end{subfigure} 
 \begin{picture}(100,0)
  \put(62,15){$\Re[\tilde{\omega}]$}
  \put(71,13){\vector(0,0){4}}
  \put(62,7){$\Im[\tilde{\omega}]$}
  \put(71,9){\vector(0,-4){4}}
  \put(0,24){$\Re [\tilde{\omega}]$}
  \put(34,22){$\Im [\tilde{\omega}]$}
  \put(98,7){$\tilde{k}$}
  \put(80,27){$\log (\tilde{\omega} \sqrt{(2 \pi n) |\tilde{d}_{*}|})$}
  \put(3,70){$\log (\tilde{k} \sqrt{(2 \pi n) |\tilde{d}_{*}|})$}
 \end{picture}
 \vskip-2em
 \caption{The spectral function (top line) and quasi-normal mode (bottom line) plots of a strongly anyonic system ($n=10^{-3}$). Uppermost figures are the logarithms of the time component of the spectral function, $\log \chi^{tt}$. Lower figures are plots of the small imaginary frequency poles of the Green's function for the same anyon density as the spectral function they are below. The solid blue lines in the lower figures are given by the analytic expression for the diffusion constant \eqref{Eq:EinsteinRelation} while the red dots are numerical data. Taking the absolute value of the red dots, $|\omega(k_{1})|$, in the lower plots gives the black dots in the spectral function plots. The black crosses in the spectral function plots are additional points obtained by a quasi-normal mode analysis not displayed in the lower plots. \textbf{Left and middle:} These figures are for $2 \pi d_{*}= -10^7$. The lower left plot is the real part of a selection of the small imaginary frequency poles while the lower middle plot is the imaginary part. \textbf{Right:} These figures are for $2 \pi \tilde{d}_{*}= - 10$. The black dots in the spectral function now only represent the pole in lower right quasi-normal mode plot with the smallest imaginary part. The lower right plot shows both real (positive value) and imaginary (negative value) parts of $\omega(k_{1})$.}
 \label{fig:PoleMotions} 
\end{figure}

% Strongly anyonised
{\ At the opposite extreme we have the strongly anyonised ($n \ll 1$) system. Surprisingly we find that the spectral function is qualitatively similar to the strongly anyonised case. As the anyon density is decreased from large to small values (top line left to right in fig.~\ref{fig:PoleMotions}) the two large peaks representing massive zero sound and diffusion join to make a single peak. This is because the pole structure follows the pattern discussed above for the weakly anyonised system. Namely, at low anyon densities two imaginary poles combine to form a complex pole at larger momentum (lower right hand plot in fig.~\ref{fig:PoleMotions}). As the anyon density is increased these poles cease to connect with each other and a distinct pole for the massive zero sound mode is formed (lower left hand plot in fig.~\ref{fig:PoleMotions}). When the momentum is increased in this case we have an imaginary pole that sinks lower into the complex plane and a separate complex pole.}

{\ There are two important facts that make the strong anyonisation regime distinct from the weak one. Firstly, the zero sound pole in the the strongly anyonised system has distinct scalings in the statistical parameter compared to the weakly anyonised zero sound mode. We shall see this in greater detail in section \ref{model:zerosound}. Secondly, for large enough anyon density, unlike in the weakly anyonised case, it is not necessary for the complex pole to have a smaller imaginary part to dominate the low frequency and momentum spectral function. This can be seen in fig.~\ref{fig:PoleMotions} by examining the plots at $n=10^{-3}$. The quasi-normal mode plots clearly show that the zero sound pole is at larger imaginary frequency than the diffusion pole and thus the diffusion pole is the relevant pole at late times. Nonetheless there is a distinct peak in the spectral function whose motion is described precisely by the zero sound quasi-normal mode.} 

%%%%%%%%%%%%%%%%%%%%%%%%%%%%%%%%%%%%%%%%%%%%%%%%%%%%%%%%%%%%%%%%%%%%%%%%%%%%%%%%%%%%%%%%%%%%%%%%%%%
\subsection{Zero sound}
\label{model:zerosound}
%%%%%%%%%%%%%%%%%%%%%%%%%%%%%%%%%%%%%%%%%%%%%%%%%%%%%%%%%%%%%%%%%%%%%%%%%%%%%%%%%%%%%%%%%%%%%%%%%%%

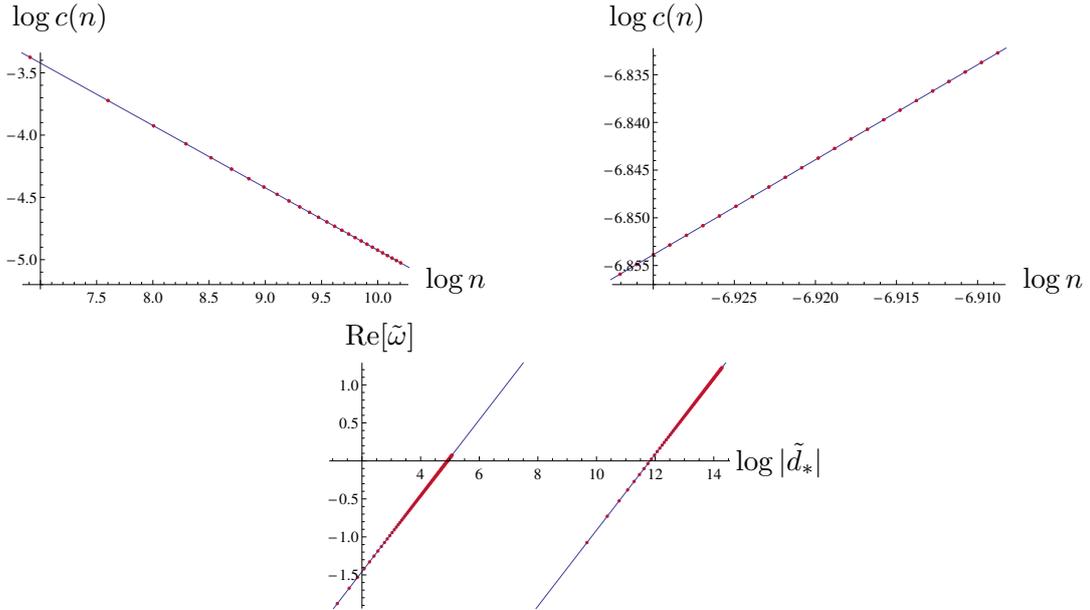
\begin{figure}[!t]
 \centering \hskip-4\unitlength 
 \begin{subfigure}
  \centering
  \includegraphics[width=0.35\textwidth]{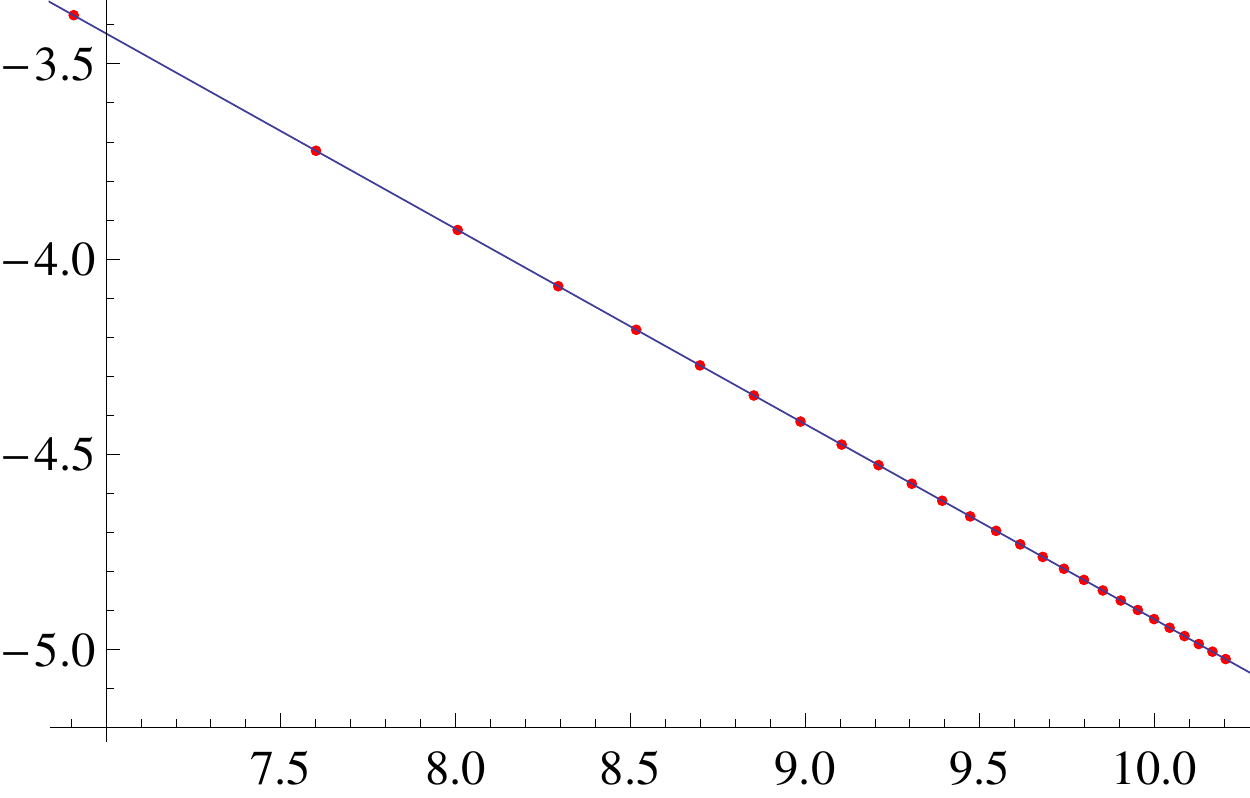}
 \end{subfigure} \qquad \qquad \qquad
 \begin{subfigure}
  \centering
  \includegraphics[width=0.35\textwidth]{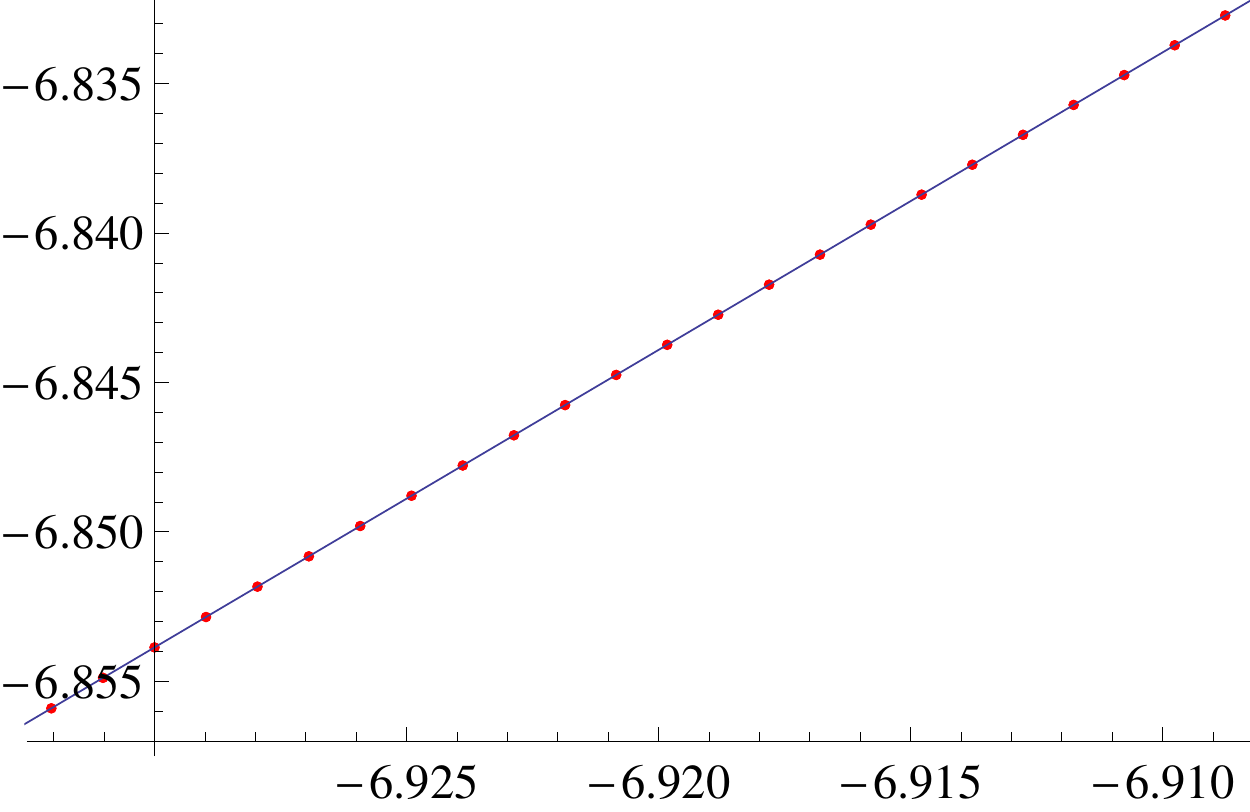}
 \end{subfigure} \\
 \begin{subfigure} 
  \centering
  \includegraphics[width=0.35\textwidth]{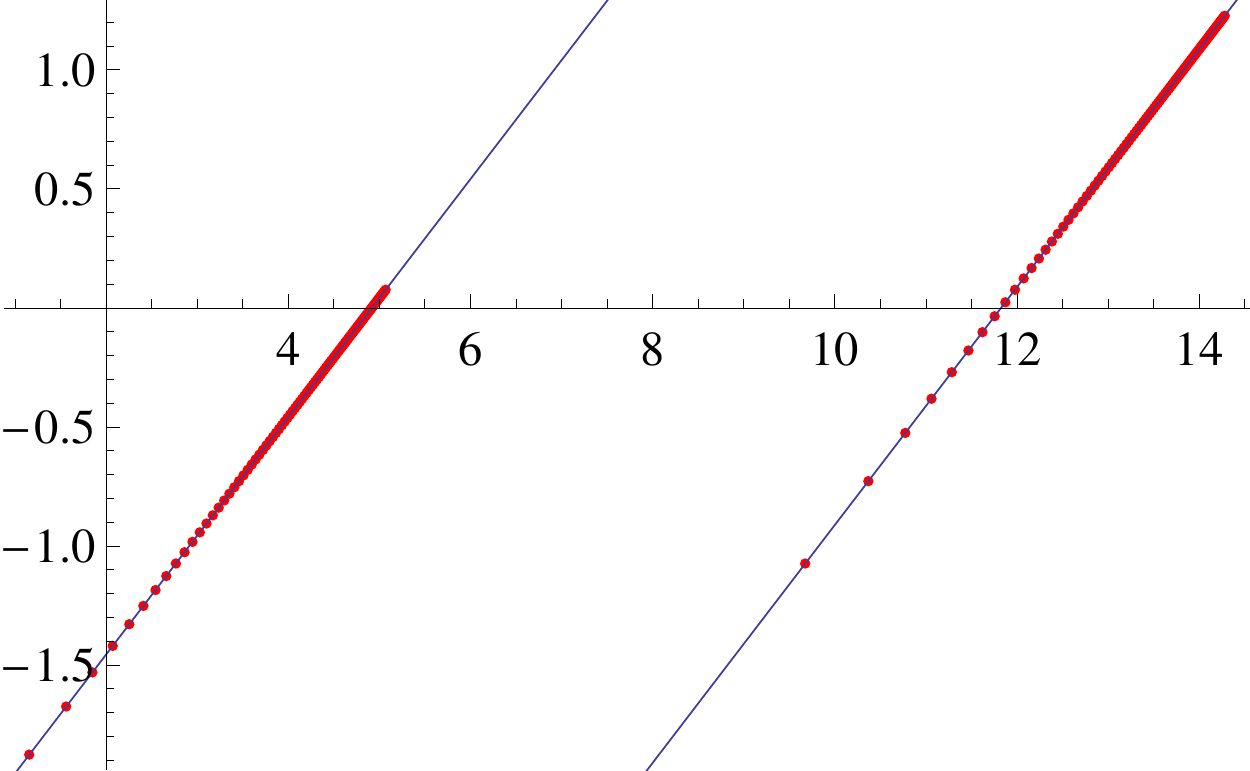}
 \end{subfigure}
 \begin{picture}(100,0)
  \put(41,33){$\log n$}
  \put(5,56){$\log c(n)$}
  \put(57,56){$\log c(n)$}
  \put(93,33){$\log n$}
  \put(34,28){$\Re[\tilde{\omega}]$}
  \put(68,17){$\log |\tilde{d}_{*}|$}
 \end{picture}
 \vskip-2em
 \caption{Figures displaying the dependence of the mass of the zero sound mode on anyon density $\tilde{d}_{*}$ and statistical parameter $n$. Solid blue lines represent best fits while the red dots are numerical data extracted from a quasi-normal mode analysis at zero momentum. \textbf{Top:} Both figures show the dependence of the zero sound mode on the statistical parameter assuming that $\tilde{\omega} = c(n) |\tilde{d}_{*}|^{1/2}$. The gradient of the best fit lines are $\approx -1/2$ (left) and $\approx +1$ (right). \textbf{Bottom:} The dependence of the mass on $\tilde{d}_{*}$ at two fixed values of the statistical parameter $n=10^3$ (i.e.~weak anyonisation) and $n=10^{-3}$ (i.e.~strong anyonisation) which are respectively the left and right lines. The gradient of both lines is approximately $1/2$. }
 \label{Fig:ZeroSound} 
\end{figure}

{\ In both the strong and weak anyonisation regimes, at sufficiently large anyon density, we find sound-like poles with non-zero masses. It is important to note that, upon complexifying the frequency of the Green's function, the poles responsible for zero sound in both regimes have distinct dependencies on the statistical parameter. Additionally for intermediate values of anyonisation it is possible for other poles in the complex plane to be closer to the real axis and thus have more of a role in the late time behaviour of the system. Taking our cue from \cite{Goykhman:2012vy,Brattan:2012nb} we fit the real part of the dispersion relations of these sound modes to an expression of the form
  \begin{eqnarray}
   \label{Eq:ZeroSound}
   \Re [\tilde{\omega}(\tilde{k}_{1})] = \pm \sqrt{\frac{1}{2} \tilde{k}_{1}^2 + c(n)^2 |\tilde{d}_{*}|} + \mathcal{O}(\tilde{k}_{1}^2)
  \end{eqnarray}
where $c(n)$s are functions of the statistical parameter which we determine using numerics. We find this to be a good description of the real part of the dispersion relations to large values of the momentum. In fig.~\ref{Fig:ZeroSound} we display the zero momentum limit of the real part of the dispersion relations against anyon density and statistical parameter. From these plots we see that
  \begin{eqnarray}
    c(n) \propto \left\{ \begin{array}{cc} n^{-1/2}, & \mathrm{weak \; anyonisation} \\ n, & \mathrm{strong \; anyonisation} \\ \end{array} \right. \; . 
  \end{eqnarray}
The weak anyonisation dependence on $n$ is commensurate with the fact that as the statistical parameter is increased the quasi-normal mode condition \eqref{Eq:QNMCondition} becomes that of the D3-D5 probe brane system prior to anyonisation. At an extremely large value of $n$ our expression for the real part of the dispersion relation gives the conformal value for the speed of sound to a good approximation. It is important to note that despite these observations the Green's function of the anyonised system will still be distinct from that of the D3-D5 probe brane system \cite{Brattan:2013wya}.}

%%%%%%%%%%%%%%%%%%%%%%%%%%%%%%%%%%%%%%%%%%%%%%%%%%%%%%%%%%%%%%%%%%%%%%%%%%%%%%%%%%%%%%%%%%%%%%%%%%%
\subsection{Diffusion and DC conductivity}
\label{model:diffusion}
%%%%%%%%%%%%%%%%%%%%%%%%%%%%%%%%%%%%%%%%%%%%%%%%%%%%%%%%%%%%%%%%%%%%%%%%%%%%%%%%%%%%%%%%%%%%%%%%%%%

\begin{figure}[t]
 \centering
 \begin{subfigure}
  \centering \hskip-6\unitlength
  \includegraphics[width=0.4\textwidth]{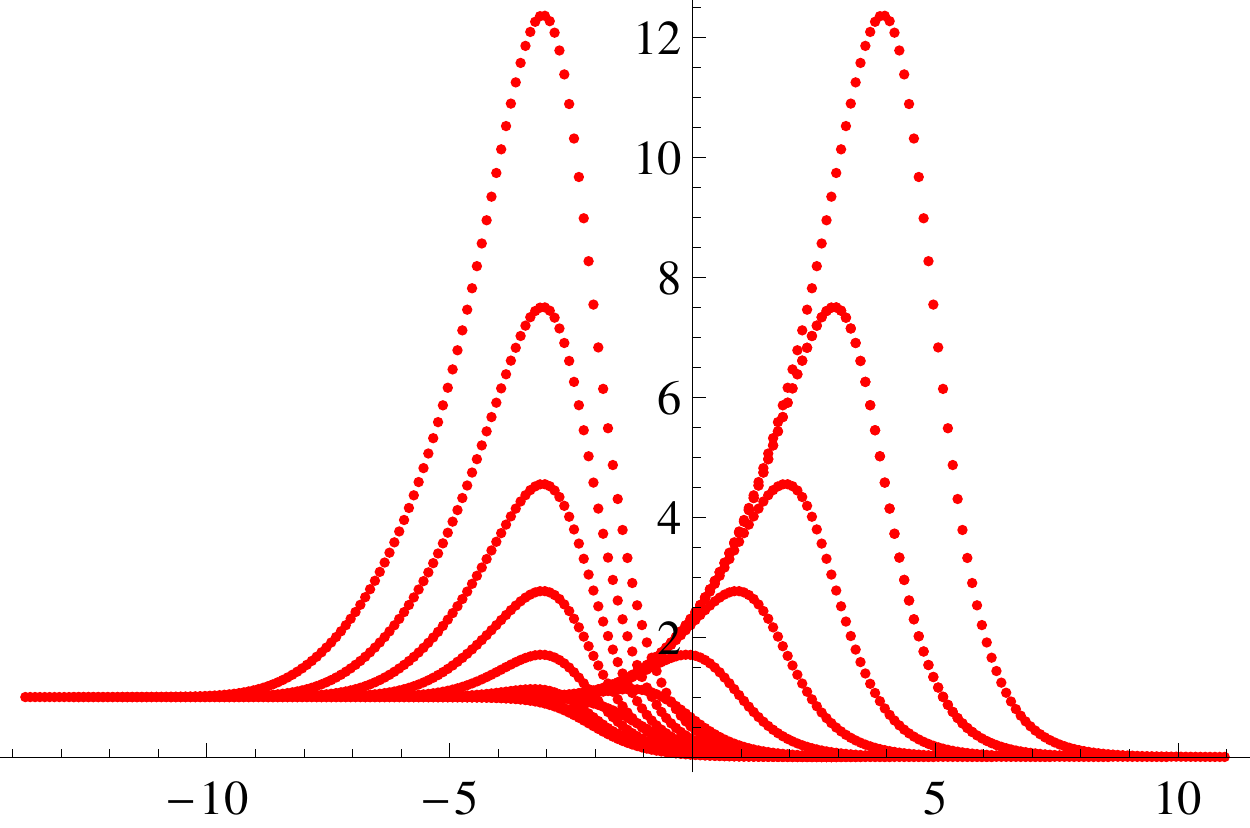}
 \end{subfigure} \hskip+8\unitlength
 \begin{subfigure}
  \centering
  \includegraphics[width=0.4\textwidth]{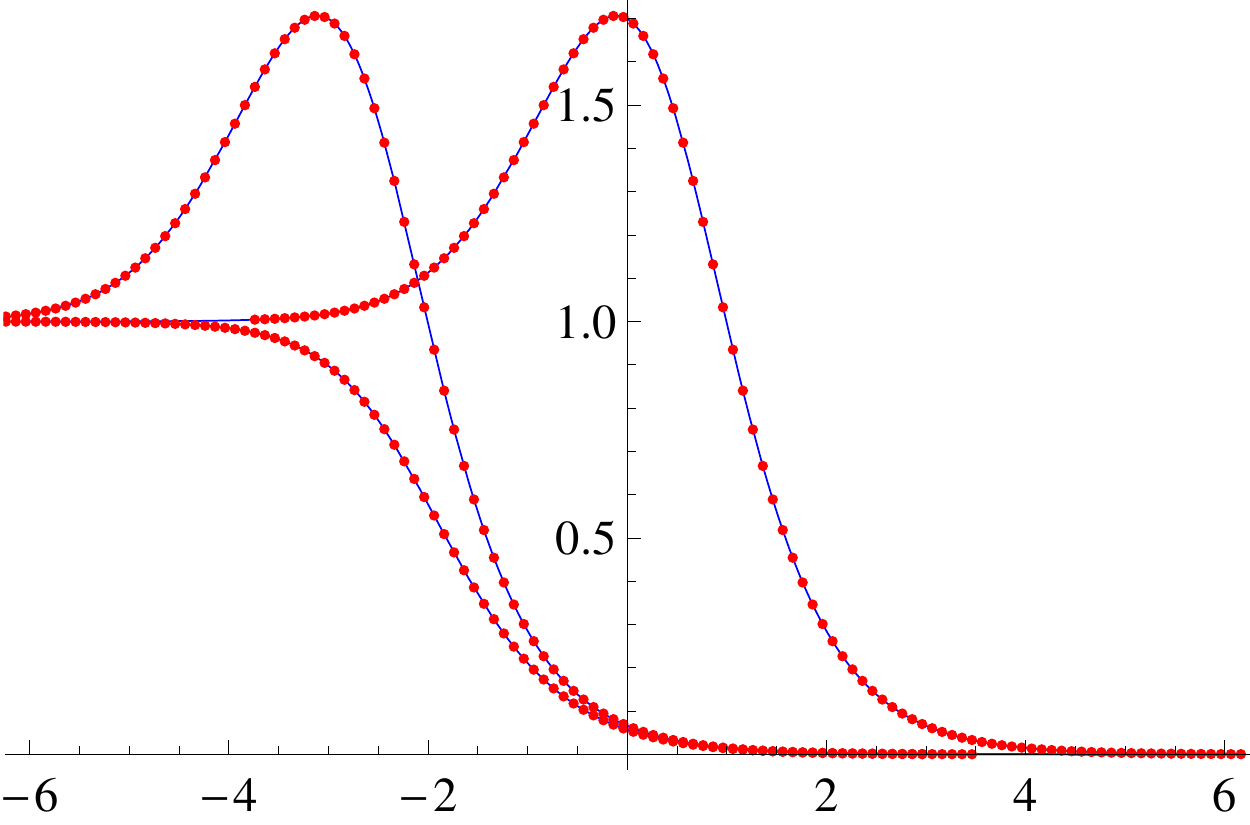}
 \end{subfigure} 
 \begin{picture}(100,0)
  \put(23,34){\small{$\tilde{D}_{*}$}}
  \put(43,7){\small{$\log |\tilde{d}_{*}|$}}
  \put(70,34){\small{$\tilde{D}_{*}$}}
  \put(93,7){\small{$\log |\tilde{d}_{*}|$}}
 \end{picture}
 \vskip-2em
 \caption{The temperature normalised diffusion constant of the anyon system against the logarithm of the anyon number density for various values of the statistical parameter. The blue lines are analytic expressions for the diffusion constant while the red dots are data points extracted from a quasi-normal mode analysis. \textbf{Left:} The diffusion constant against $\log \tilde{d}_{*}$ for $\log n = -7, -6 , \ldots, 6, 7$. The largest peaks correspond to the largest absolute values of $\log(n)$ with $n\ll 1$ constituting the right hand peaks whose position increases with $n$ and $n \gg 1$ being the left hand peaks whose position is approximately constant. \textbf{Right:} A comparison of the Einstein relation given by \eqref{Eq:EinsteinRelation} and the quasi-normal mode data against $\log \tilde{d}_{*}$ for $\log n=-3$ (leftmost peak), $\log n = 0$ (no peak) and $\log n = 3$ (rightmost peak).}
 \label{fig:DiffusionConstants}
\end{figure}
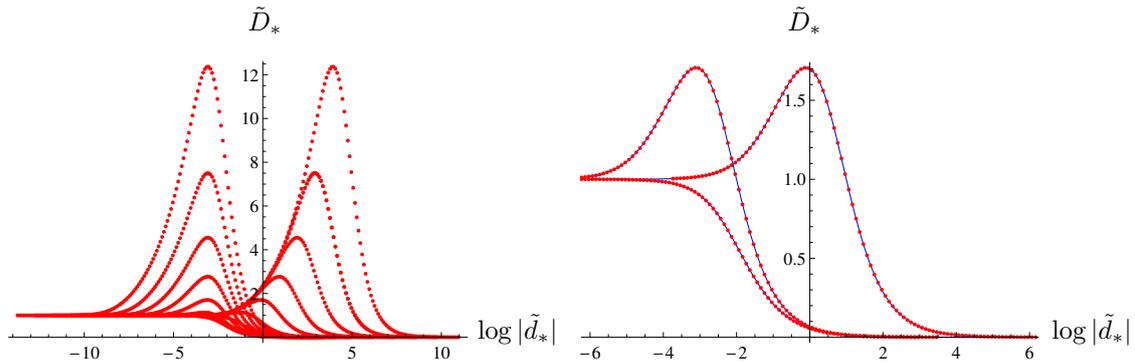

% Describe the general features of the graph
{\ In fig.~\ref{fig:DiffusionConstants} we display the diffusion constant for various anyon densities and statistical parameters. The value of the temperature normalised diffusion constant interpolates between one at small $|\tilde{d}_{*}|$, which is a fixed multiple of the conformal value, and zero at large $|\tilde{d}_{*}|$ for all $n$. Picking a value of $n$ and increasing $|\tilde{d}_{*}|$ one finds that initially the diffusion constant increases before peaking and beginning to decrease for most values of $n$. There is a range of $n$ however, including $n=1$, where this is not true and there is a monotonic interpolation between the values of one and zero i.e.~no peak.}

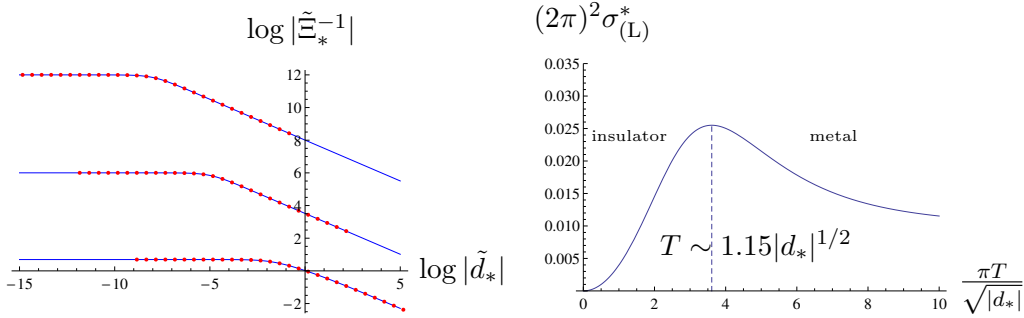
\begin{figure}[t]
 \centering \hskip-6\unitlength
 \begin{subfigure}
  \centering
  \includegraphics[width=0.35\textwidth]{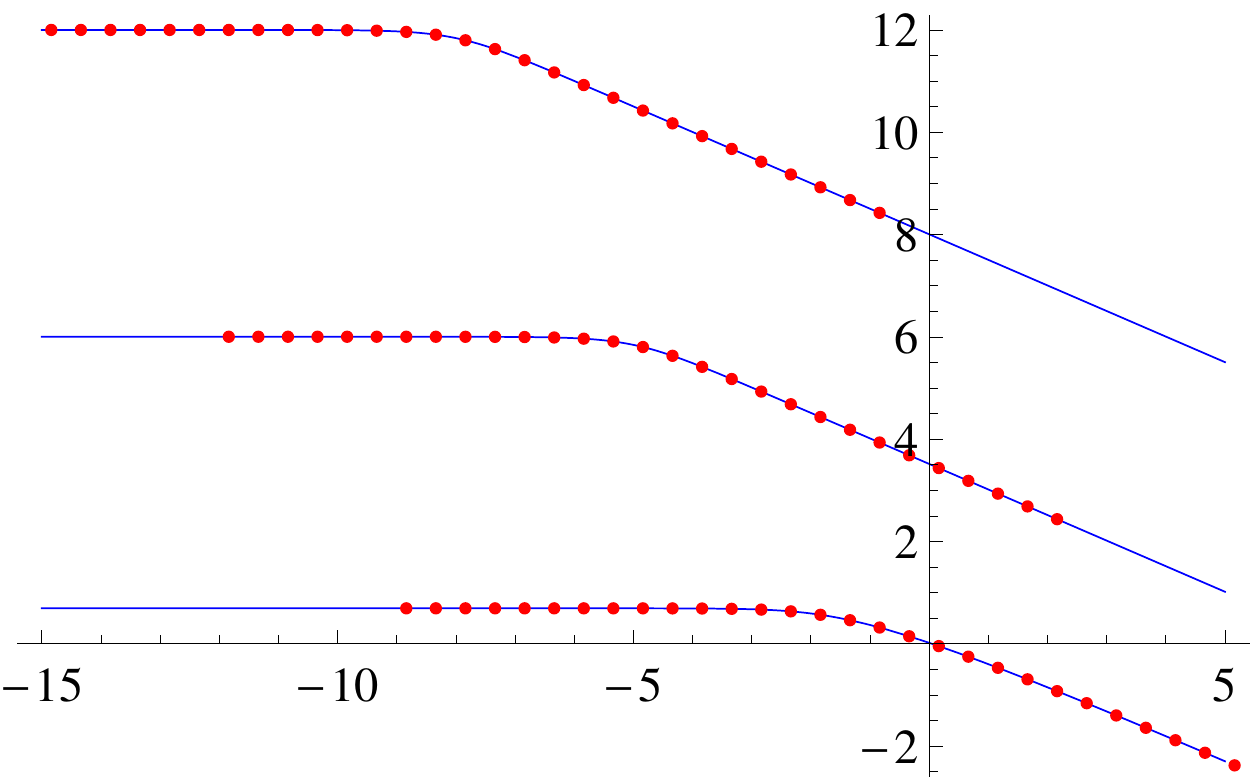}
 \end{subfigure} \qquad \qquad
 \begin{subfigure}
  \centering
  \includegraphics[width=0.35\textwidth]{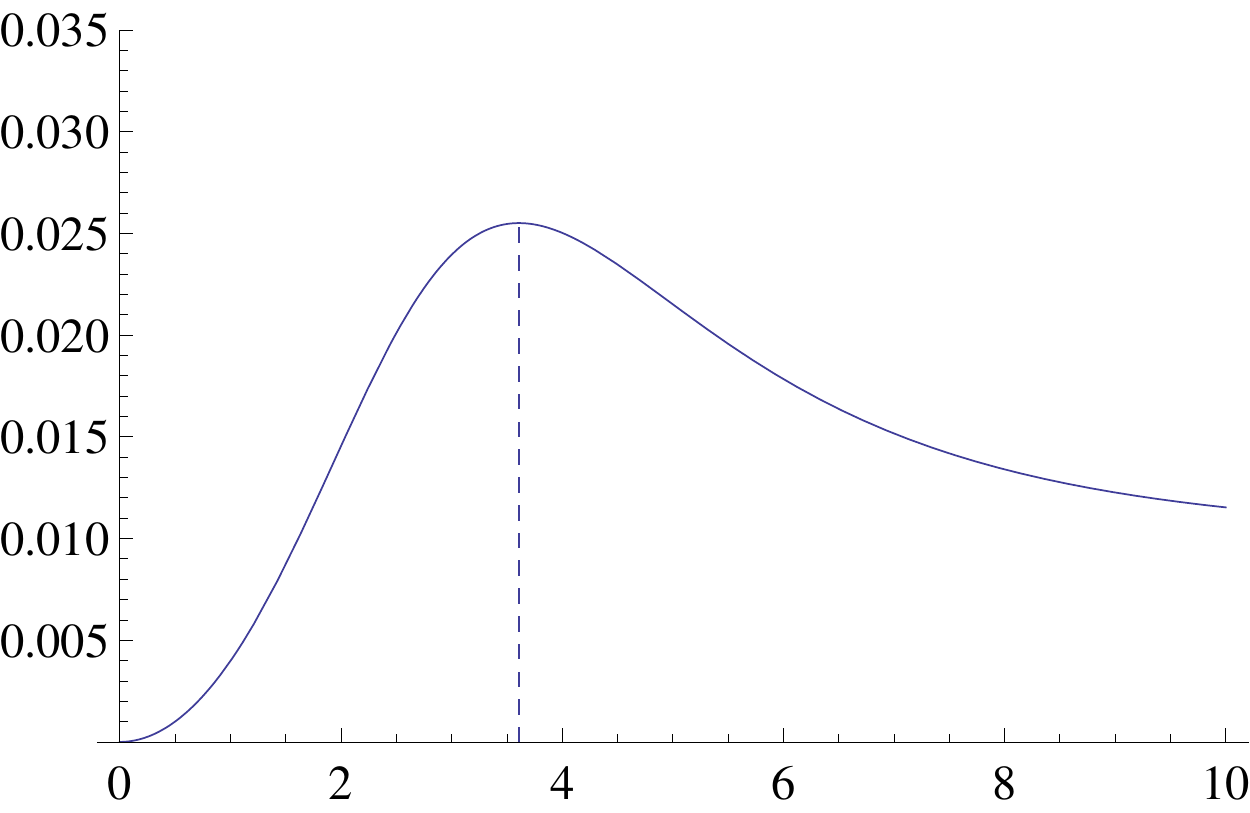}
 \end{subfigure}
 \begin{picture}(100,0)
  \put(27,29){$\log |\tilde{\Xi}_{*}^{-1}|$}
  \put(42,8){$\log |\tilde{d}_{*}|$}
  \put(52,30){$(2\pi)^2 \sigma_{(\mathrm{L})}^{*}$}
  \put(63,10){$T \sim 1.15 |d_{*}|^{1/2}$}
  \put(76,20){\tiny{metal}}
  \put(57,20){\tiny{insulator}}
  \put(89,7){$\frac{\pi T}{\sqrt{|d_{*}|}}$}
 \end{picture}
 \vskip-2em
 \caption{\textbf{Left:} The inverse charge susceptibility against $\log |\tilde{d}_{*}|$ for various values of the statistical parameter - $\log n = 0, 3, 6$. The blue line is the analytic result from \eqref{Eq:AnyonSusceptilibility} while red dots are numerical points extracted from the quasi-normal mode analysis. \textbf{Right:} The longitudinal conductivity at $n=10$ (solid blue line) against the temperature normalised by density. We see a critical temperature indicated by the blue dashed line that depends on the square root of the anyon density. The overall numerical factor determining the position of the critical temperature is a function of $n$ and is given by \eqref{Eq:Textrema}. The metallic and insulating regions are indicated by the labels. We see at $T=0$ there is no longitudinal conductivity and the material is an insulator in the longitudinal directions.}
 \label{fig:Susceptibilities}
\end{figure}

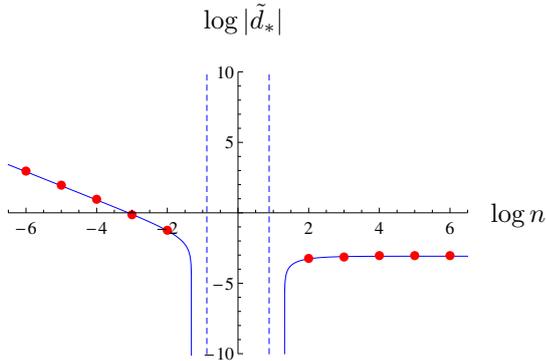
\begin{figure}[t]
 \centering{}
 \includegraphics[width=0.4\textwidth]{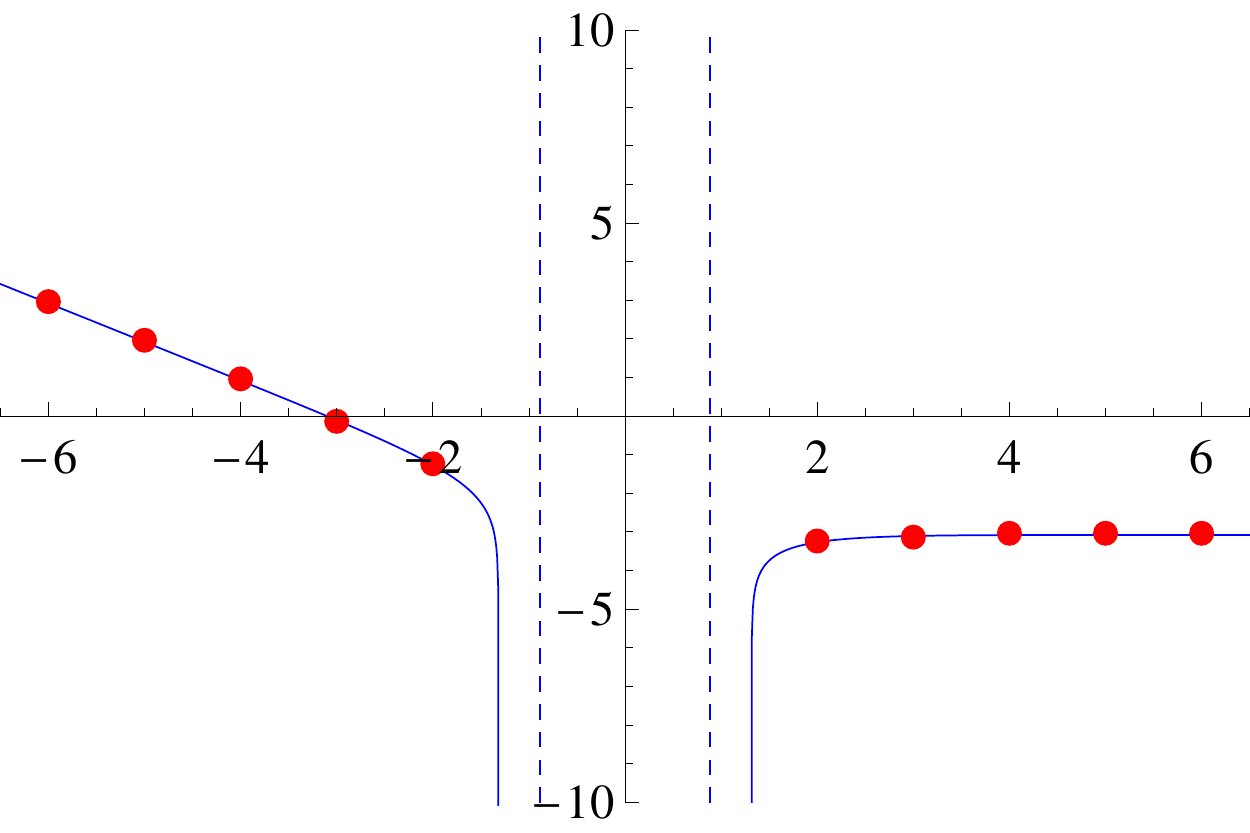}
 \begin{picture}(100,0)
  \put(47,33){\small{$\log |\tilde{d}_{*}|$}}
  \put(72,16){\small{$\log n$}}
 \end{picture}
 \vskip-2em
 \caption{The value of $\log |\tilde{d}_{*}|$ corresponding to the diffusion constant peak against $\log(n)$. The red dots are numerical data extracted from a quasinormal mode analysis while the solid blue line is the numerical solution to minimizing \eqref{Eq:EinsteinRelation} with respect to $\tilde{d}_{*}$. The vertical blue dashed lines represent bounds on when the longitudinal DC conductivity has a peak. As the value of $|d_{*}|$ increases we can see that the maximum value of the diffusion constant occurs for a fixed value of $\tilde{d}_{*} \approx -0.05$.}
 \label{fig:MaxDiffusionConstant}
\end{figure}

% Einstein relation and DC conductivities
{\ The behaviour of the diffusion constant can be understood by considering the Einstein relation which relates charge susceptibility, longitudinal DC conductivity and the diffusion constant
  \begin{eqnarray}
    \label{Eq:EinsteinRelation}
    \tilde{D}_{*} = \frac{\tilde{\sigma}_{(\mathrm{L})}^{*}}{\tilde{\Xi}_{*}} \; . 
  \end{eqnarray}
To obtain the charge susceptibility we note that for sufficiently small values of the momentum the retarded Green's function should reproduce the thermodynamic susceptibilities. In particular, for the time component, we have
  \begin{eqnarray}
   [\tilde{G}_{R}^{*}(0,\tilde{k})]^{tt} &\stackrel{\tilde{k} \ll 1}{=}& \left( \frac{\partial \tilde{d}_{*}}{\partial \tilde{\mu}_{*}} \right)_{\tilde{B}_{*}} \; . 
  \end{eqnarray}
This can be computed by performing the appropriate $SL(2,\mathbb{Z})$ transformation of the current-current correlator of the D3-D5 probe brane system (see appendix \ref{chargesusceptibility}). Alternatively it can be extracted from the numerical Green's function of the anyon system. We compare both of these approaches in fig.~\ref{fig:Susceptibilities}. We see clearly that the charge susceptibility is a monotonic function of the anyon density. The DC conductivities can readily be computed analytically and have the form
  \begin{eqnarray}
   \label{Eq:DCconductivities}
   \sigma^{*}_{(\mathrm{L})} = \frac{1}{(2\pi)^2} \frac{\sqrt{1 + (2 \pi) ^2 \tilde{d}_{*}^2 \left(1+n^2\right)}}{1+ \left(1+ 4 (2 \pi)^2 \tilde{d}_{*}^2 \right) n^2} \; , \qquad 
   \sigma^{*}_{(\mathrm{H})} = \frac{n}{(2\pi)^2} \frac{1 + 2 \left( 2 \pi \tilde{d}_{*} \right)^2}{1+ \left(1 + 4 (2 \pi)^2 \tilde{d}_{*}^2\right) n^2} \; . \qquad
  \end{eqnarray}
The rightmost plot in fig.~\ref{fig:DiffusionConstants} shows that the Einstein relation is indeed giving precisely the value of the diffusion constant. To find the extrema of the diffusion constant requires minimizing \eqref{Eq:EinsteinRelation} which we could not do analytically. Instead we solved it numerically and compared to the maxima extracted from the quasi-normal mode analysis as displayed in fig.~\ref{fig:MaxDiffusionConstant}.}  

{\ As was noted previously, the charge susceptibility is monotonic in $\tilde{d}_{*}$. As such the existence or not of peaks in the diffusion constant are due to a feature in the DC conductivities - namely the metal-insulator transition. In fig.~\ref{fig:Susceptibilities} we plot the longitudinal conductivity as a function of temperature. We see that as the temperature goes to zero the longitudinal conductivity vanishes as $T^2$. Hence we identify the material as an insulator. It may be tempting to identify it as a perfect insulator given that $\sigma_{({L})}^{*}(0) \equiv 0$ however we note that $\sigma_{(\mathrm{H})}^{*}(0) = 1/(2n)$ and thus our anyon material still has a non-zero component in it's conductivity tensor at zero temperature. At large temperatures the conductivities asymptote to $1/(1+n^2)$ and $n/(1+n^2)$ respectively. Also in fig.~\ref{fig:Susceptibilities}, because we have chosen an appropriate $n$, there is a peak in the longitudinal DC conductivity. More generally the critical temperature 
at which the longitudinal conductivity has turning points is given by
  \begin{eqnarray} 
   \label{Eq:Textrema}
   T_{\mathrm{critical}}  = \frac{2 \sqrt{n} (1+n^2)^{1/4}}{\sqrt{\pi} (n^4 - 6 n^2 + 1)^{1/4}} |d_{*}|^{1/2} \; . 
  \end{eqnarray}
Given that the temperature is real and positive the critical temperature only exists for $0 \leq n<\sqrt{2}-1$ and $n>1+\sqrt{2}$. Between these bounds the peak vanishes and there is a smooth interpolation between insulating and conducting behaviour. We note that there is no such peak in the DC Hall conductivity and it smoothly interpolates between its zero and large temperature bounds.}

% Generic features
{\ Now that we have discussed the analytic expression for the diffusion constant we can explain the features of fig.~\ref{fig:DiffusionConstants} in more detail. We note first that for small anyon density the diffusion constant tends to one in our units, which is the conformal value. For most choices of the statistical parameter increasing the anyon number density then causes the diffusion constant to increase. This reflects the fact that initially as the number of carriers increases it is easier for number density fluctuations to diffuse. On the other hand the diffusion constant always tends to zero at large anyon density. This behaviour is due to the effect of non-standard statistics. Thus a lump of additional anyon density when introduced into a sufficiently dense system holds together for a long time. The anyons have more attraction to each other than the bosons and fermions from which they are made. This is the interpretation of the peaks in the diffusion constant, and the fact that the diffusion constant falls to zero, for fixed $n$ and increasing density. Finally we note that for a window about $n \sim 1$, as displayed in fig.~\ref{fig:MaxDiffusionConstant}, there are no peaks in the diffusion constant. In this case there is a unique balancing between the attraction of the anyons to each other and the benefit of having more charge carriers to diffuse perturbations.}

%%%%%%%%%%%%%%%%%%%%%%%%%%%%%%%%%%%%%%%%%%%%%%%%%%%%%%%%%%%%%%%%%%%%%%%%%%%%%%%%%%%%%%%%%%%%%%%%%%%
\subsection{AC conductivity}
\label{model:ACconductivity}
%%%%%%%%%%%%%%%%%%%%%%%%%%%%%%%%%%%%%%%%%%%%%%%%%%%%%%%%%%%%%%%%%%%%%%%%%%%%%%%%%%%%%%%%%%%%%%%%%%%

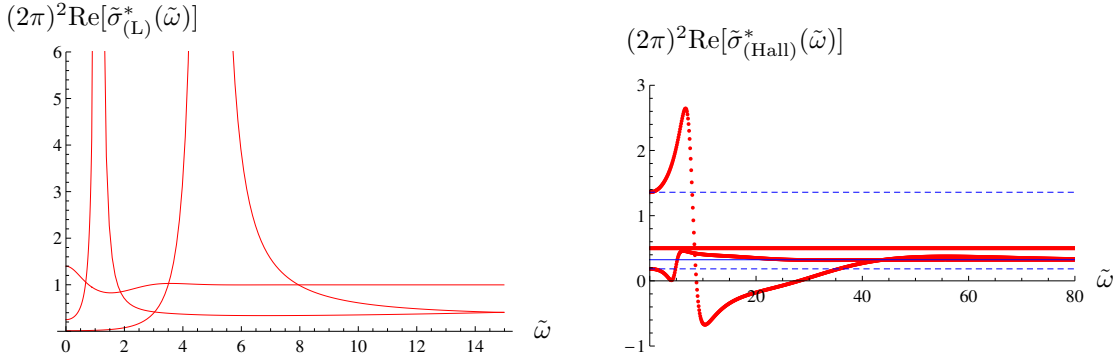
\begin{figure}[t]
 \centering
 \begin{subfigure}
  \centering \hskip-2\unitlength
  \includegraphics[width=0.4\textwidth]{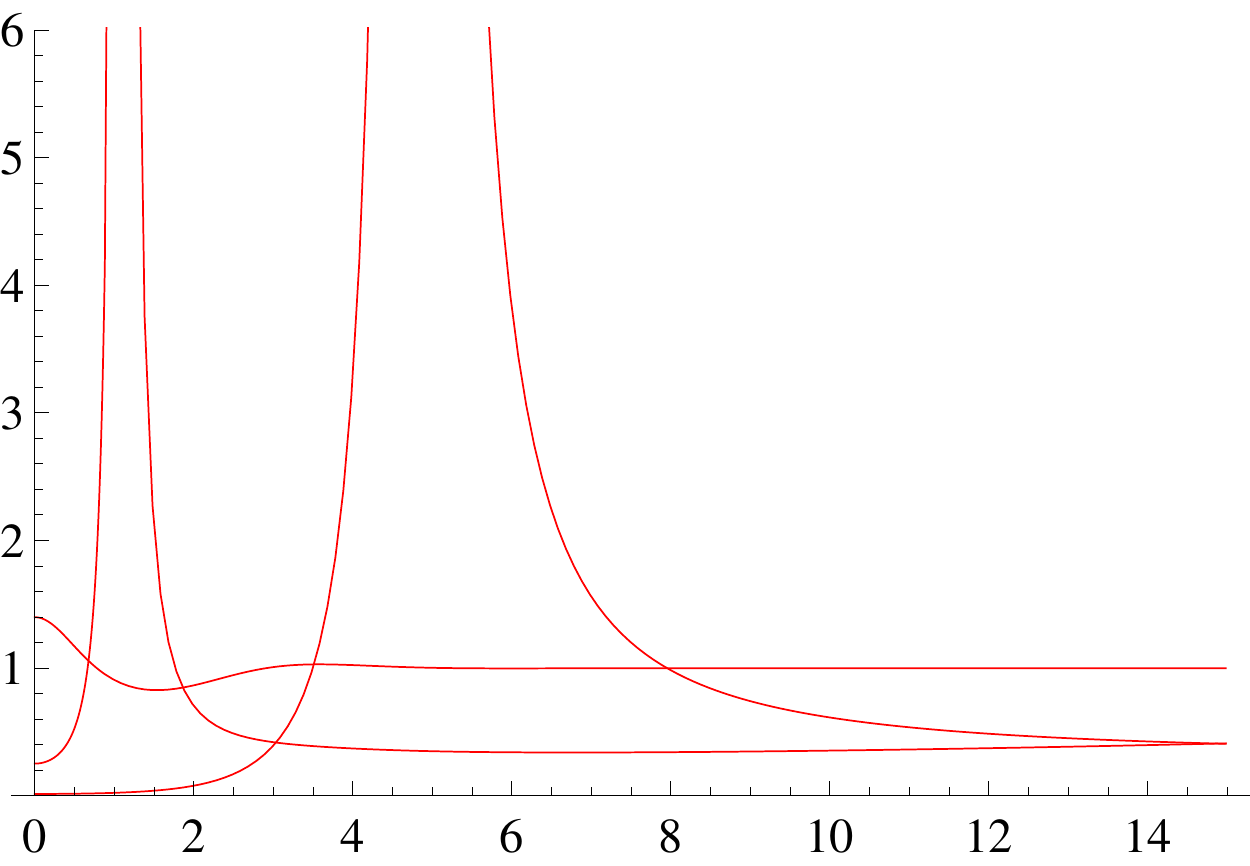}
 \end{subfigure} \hskip+8\unitlength
 \begin{subfigure}
  \centering
  \includegraphics[width=0.4\textwidth]{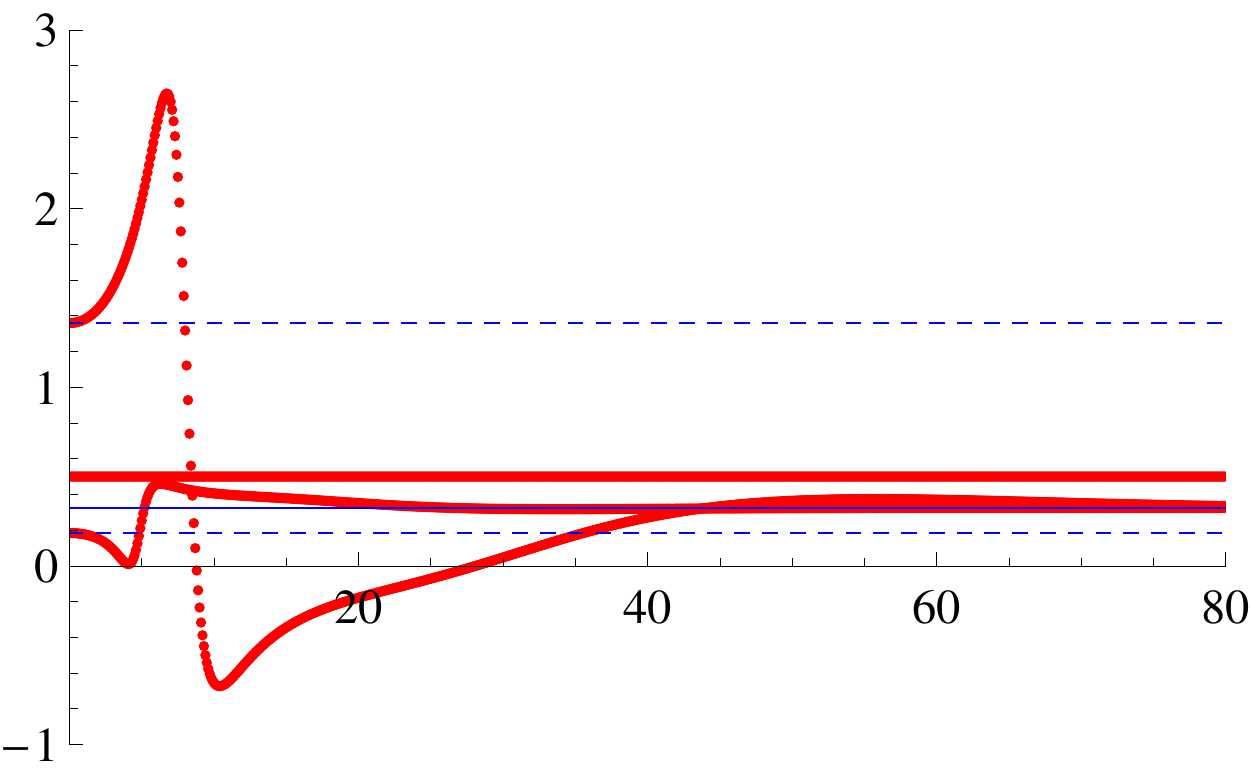}
 \end{subfigure} 
 \begin{picture}(100,0)
  \put(0,34){\small{$(2\pi)^2 \Re[\tilde{\sigma}^{*}_{\mathrm{(L)}}(\tilde{\omega})]$}}
  \put(46,6.5){\small{$\tilde{\omega}$}}
  \put(54,32){\small{$(2\pi)^2 \Re[\tilde{\sigma}^{*}_{\mathrm{(Hall)}}(\tilde{\omega})]$}}
  \put(95,11){\small{$\tilde{\omega}$}}
 \end{picture}
 \vskip-2em
 \caption{The real parts of the AC longitudinal and Hall conductivities against frequency for various choices of density and statistical parameter. \textbf{Left:} The AC longitudinal conductivity for $\log n=-3$ and $\log (- 2 \pi \tilde{d}_{*})=0,6,9$ where the solid red line is an interpolation of data extracted from the Green's function. We can see the change from a Drude-like behaviour ($\log (- 2 \pi \tilde{d}_{*})=0$, smallest peak) to incoherent metal ($\log  (- 2 \pi \tilde{d}_{*})=6$, first displaced peak) to insulator ($\log (- 2 \pi \tilde{d}_{*})=9$, approximately zero at $\tilde{\omega}=0$). \textbf{Right:} The AC Hall conductivity various anyon densities and statistical parameters. The red dots are data, the solid blue line is the zero density value of the DC Hall conductivity and the dashed blue line is the DC value of the Hall conductivity at the given values of $n$ and $\tilde{d}_{*}$. The red points with the largest range belong to $\log n =-1,\log (- 2\pi \tilde{d}_{*})=6$ while the flat red line is the value of the AC Hall conductivity at $\log n = 0,\log (- 2\pi \tilde{d}_{*})=5$. The remaining line is for $n=1, \log (- 2\pi \tilde{d}_{*})=4$.}
 \label{fig:ACconductivities}
\end{figure}

% What are the features in the AC longitudinal conductivity
{\ The longitudinal conductivity at small anyon densities is peaked at $\tilde{\omega}=0$ and, for all values of the anyon density, decays to the conformal value of one at large $\tilde{\omega}$. At larger values of $|\tilde{d}_{*}|$ the peak becomes displaced to some non-zero value of $\tilde{\omega}$. Increasing the density still further causes the DC longitudinal conductivity to drop to very small values. The anyon system moves as the density is increased from being a Drude-like conductor (by which we mean there is a major peak at $\tilde{\omega}=0$), through incoherent metal behaviour (where the major peak is displaced to non-zero $\tilde{\omega}$) to an insulator (the DC conductivity drops to very small values).}

% AC hall conductivity
{\ The AC Hall conductivity interpolates between the DC Hall conductivity \eqref{Eq:DCconductivities} and its zero density limit $n/(1+n^2)$. At $\tilde{\omega} \sim 0$ it takes the DC value but as the frequency increases there is a peak and the value rapidly drops to the zero density limit.  If $\log n > 0$ the zero density limit is smaller than the finite density value while it is the other way around if $\log n < 0$. Three examples of the AC Hall conductivity are displayed in fig.~\ref{fig:ACconductivities} where this behaviour can clearly be recognised. When $\log n = 0$ the AC Hall conductivity is constant and has the value one. This is consistent with the above observations as when $n=1$ the expression of \eqref{Eq:DCconductivities} reduces to $1/2$. The length of the initial plateau where $\sigma_{\mathrm{Hall}}^{*}(\omega)$ is approximately constant can be increased by increasing the anyon density. This feature is generic in the regime investigated.}

%%%%%%%%%%%%%%%%%%%%%%%%%%%%%%%%%%%%%%%%%%%%%%%%%%%%%%%%%%%%%%%%%%%%%%%%%%%%%%%%%%%%%%%%%%%%%%%%%%%
%%%%%%%%%%%%%%%%%%%%%%%%%%%%%%%%%%%%%%%%%%%%%%%%%%%%%%%%%%%%%%%%%%%%%%%%%%%%%%%%%%%%%%%%%%%%%%%%%%%
%%%%%%%%%%%%%%%%%%%%%%%%%%%%%%%%%%%%%%%%%%%%%%%%%%%%%%%%%%%%%%%%%%%%%%%%%%%%%%%%%%%%%%%%%%%%%%%%%%%
\section{Future directions}
\label{future}
%%%%%%%%%%%%%%%%%%%%%%%%%%%%%%%%%%%%%%%%%%%%%%%%%%%%%%%%%%%%%%%%%%%%%%%%%%%%%%%%%%%%%%%%%%%%%%%%%%%
%%%%%%%%%%%%%%%%%%%%%%%%%%%%%%%%%%%%%%%%%%%%%%%%%%%%%%%%%%%%%%%%%%%%%%%%%%%%%%%%%%%%%%%%%%%%%%%%%%%
%%%%%%%%%%%%%%%%%%%%%%%%%%%%%%%%%%%%%%%%%%%%%%%%%%%%%%%%%%%%%%%%%%%%%%%%%%%%%%%%%%%%%%%%%%%%%%%%%%%

{\ In this paper we have shown that many of the qualitative phenomena associated with non-zero magnetic fields in the finite density D3-D5 probe system, such as the existence of massive zero sound and plateaus in the Hall conductivity, are mirrored in the anyon system. The most important difference however is that these features occur in the anyon system in the absence of a magnetic field. We have also demonstrated evidence for an analogue of the metal-insulator transition for strongly coupled anyons.}

{\ Another phase which can potentially coexist with the one discussed here is given by turning on one of the embedding scalars. In the original D3-D5 system this corresponds to breaking the chiral symmetry of the fermions in the strongly coupled field theory \cite{Evans:2010iy,Evans:2010hi,Pal:2010gj,Jensen:2010ga,Evans:2010np}. While it would be interesting to understand the effect that chiral symmetry breaking has on anyon physics in its own right it would also be interesting to understand when the ground state considered here is dominant. We would generically expect some difference to the D3-D5 probe brane case as the additional boundary terms do make a contribution to the on-shell action. We leave this for future research.}

\acknowledgments

{\ DB is supported by in part by the Israeli Science Foundation (ISF) 392/09 and a Fine Fellowship. DB would like to thank Oren Bergman, Niko Jokela, Gilad Lifschytz, Matthew Lippert, Andy O'Bannon, Alfonso V\'{a}zquez Ramallo and Amos Yarom for useful discussions.}

\appendix

%%%%%%%%%%%%%%%%%%%%%%%%%%%%%%%%%%%%%%%%%%%%%%%%%%%%%%%%%%%%%%%%%%%%%%%%%%%%%%%%%%%%%%%%%%%%%%%%%%%
%%%%%%%%%%%%%%%%%%%%%%%%%%%%%%%%%%%%%%%%%%%%%%%%%%%%%%%%%%%%%%%%%%%%%%%%%%%%%%%%%%%%%%%%%%%%%%%%%%%
%%%%%%%%%%%%%%%%%%%%%%%%%%%%%%%%%%%%%%%%%%%%%%%%%%%%%%%%%%%%%%%%%%%%%%%%%%%%%%%%%%%%%%%%%%%%%%%%%%%
\section{Charge susceptibility by $SL(2,\mathbb{Z})$}
\label{chargesusceptibility}
%%%%%%%%%%%%%%%%%%%%%%%%%%%%%%%%%%%%%%%%%%%%%%%%%%%%%%%%%%%%%%%%%%%%%%%%%%%%%%%%%%%%%%%%%%%%%%%%%%%
%%%%%%%%%%%%%%%%%%%%%%%%%%%%%%%%%%%%%%%%%%%%%%%%%%%%%%%%%%%%%%%%%%%%%%%%%%%%%%%%%%%%%%%%%%%%%%%%%%%
%%%%%%%%%%%%%%%%%%%%%%%%%%%%%%%%%%%%%%%%%%%%%%%%%%%%%%%%%%%%%%%%%%%%%%%%%%%%%%%%%%%%%%%%%%%%%%%%%%%

{\ The current-current correlator for the D3-D5 probe brane system with total charge $Q= \tilde{d} (\pi T)^2 \mathcal{N}_{5} V$ and magnetic field $B=\tilde{B} (\pi T)^2$ at temperature $T$ in the low momentum limit is
  \begin{eqnarray}
   \tilde{G}_{R}(0,\tilde{k}) &\stackrel{\tilde{k} \ll 1}{=}& \left(
								    \begin{array}{cc}
								      \left( \frac{\partial \tilde{d}}{\partial \tilde{\mu}} \right)_{\tilde{B}} & \tilde{k} \left( \frac{\partial \tilde{d}}{\partial \tilde{B}} \right)_{\tilde{\mu}} \\
								      \tilde{k} \left( \frac{\partial \tilde{m}}{\partial \tilde{\mu}} \right)_{\tilde{B}_{*}} & \tilde{k}^2 \left( \frac{\partial \tilde{m}}{\partial \tilde{B}} \right)_{\tilde{\mu}} \\
								    \end{array}
								 \right) \; ,
  \end{eqnarray}
where
  \begin{eqnarray}
    \tilde{\mu}(\tilde{d},\tilde{B}) &=& \frac{\tilde{d} \left(\frac{\Gamma \left(\frac{1}{4}\right) \Gamma \left(\frac{5}{4}\right) \sqrt[4]{\tilde{d}^2+\tilde{B}}}{\sqrt{\pi }}-\,
   _2F_1\left(\frac{1}{4},\frac{1}{2};\frac{5}{4};-\frac{1}{\tilde{d}^2+\tilde{B}^2}\right)\right)}{\sqrt{\tilde{d}^2+\tilde{B}^2}} \; , \\
    \tilde{m}(\tilde{d},\tilde{B}) &=& \tilde{B} \, _2F_1\left(\frac{1}{4},\frac{1}{2};\frac{5}{4};-\tilde{B}^2-\tilde{d}^2\right) \; .
  \end{eqnarray}
The charge susceptibility of the anyon system is then
  \begin{eqnarray}
       \label{Eq:AnyonSusceptilibility}
       \left( \frac{\partial \tilde{d}_{*}}{\partial \tilde{\mu}_{*}} \right)_{B_{*}}
   &=& \frac{\left(\frac{\partial \tilde{d}}{\partial \tilde{\mu}} \right)_{\tilde{B}}}{\left( \frac{\partial \tilde{d}}{\partial \tilde{\mu}} \right)_{\tilde{B}} \left( \frac{\partial \tilde{m}}{\partial \tilde{B}} \right)_{\tilde{\mu}} + \left(\left( \frac{\partial \tilde{d}}{\partial \tilde{B}} \right)_{\tilde{\mu}} - n \right)^2 }
  \end{eqnarray}
where we must take $\tilde{d} = - 2 \pi n \tilde{d}_{*}$ and $\tilde{B} = - 2 \pi \tilde{d}_{*}$.}
  
\bibliographystyle{JHEP}
\bibliography{alt_quant}

\end{document}